\documentclass[twocolumn]{aastex63}
\usepackage{graphicx}
\usepackage{amssymb}
\usepackage{amsmath}
\usepackage{wasysym}
\usepackage{xfrac}
\usepackage{float}

\newcommand{\DIW}{$\Delta$IW }
\accepted{ApJ, August 8, 2020}
\shortauthors{Doyle et al.}
\shorttitle{Where are the Extrasolar Mercuries?}

\begin{document}

\title{Where are the Extrasolar Mercuries?}

\correspondingauthor{Alexandra E. Doyle}
\email{a.doyle@ucla.edu}

\correspondingauthor{Edward D. Young}
\email{eyoung@epss.ucla.edu}

\author{Alexandra E. Doyle}
\affiliation{Earth, Planetary, and Space Sciences\\
University of California, Los Angeles\\
Los Angeles, CA 90095, USA}

\author{Beth Klein}
\affiliation{Physics and Astronomy\\
University of California, Los Angeles\\
Los Angeles, CA 90095, USA}

\author{Hilke E. Schlichting}
\affiliation{Earth, Planetary, and Space Sciences\\
University of California, Los Angeles\\
Los Angeles, CA 90095, USA}
\affiliation{Physics and Astronomy\\
University of California, Los Angeles\\
Los Angeles, CA 90095, USA}
\affiliation{Earth, Atmospheric and Planetary Sciences\\
Massachusetts Institute of Technology\\
Cambridge, MA 02139, USA}

\author{Edward D. Young}
\affiliation{Earth, Planetary, and Space Sciences\\
University of California, Los Angeles\\
Los Angeles, CA 90095, USA}

\begin{abstract}

We utilize observations of 16 white dwarf stars to calculate and analyze the oxidation states of the parent bodies accreting onto the stars. Oxygen fugacity, a measure of overall oxidation state for rocks, is as important as pressure and temperature in determining the structure of a planet. We find that most of the extrasolar rocky bodies formed under oxidizing conditions, but $\thicksim$1/4 of the polluted white dwarfs have compositions consistent with more reduced parent bodies. The difficulty in constraining the oxidation states of relatively reduced bodies is discussed and a model for the time-dependent evolution of the apparent oxygen fugacity for a hypothetical reduced body engulfed by a WD is investigated. Differences in diffusive fluxes of various elements through the WD envelope yield spurious inferred bulk elemental compositions and oxidation states of the accreting parent bodies under certain conditions. The worst case for biasing against detection of reduced bodies occurs for high effective temperatures. For moderate and low effective temperatures, evidence for relatively reduced parent bodies is preserved under most circumstances for at least several characteristic lifetimes of the debris disk.

\end{abstract}

\section{Introduction}

\begin{figure*}
\begin{center}
\includegraphics[width=5.5in]{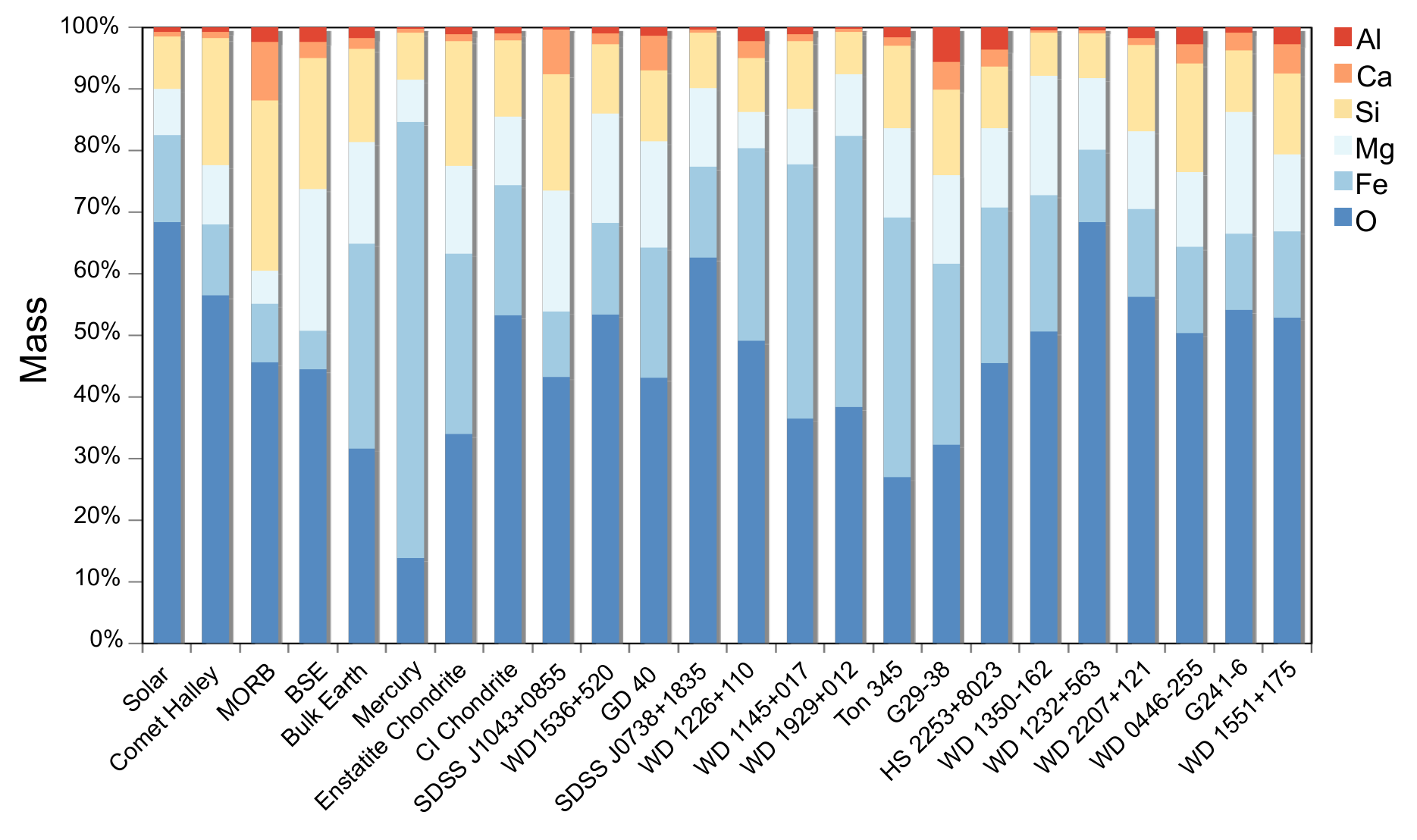} 
\caption{Mass concentrations of the 6 major rock-forming elements in the 16 WDs used in this study normalized to 100\%. Generally, rocky bodies accreting onto these stars resemble rocks in our own Solar System and range from basaltic, crust-like to bulk-asteroid or planet-like compositions. The excess oxygen seen in WD 1232+563 is discussed in Section \ref{SettlingModel}. Note WD 1145+017 is updated from \cite{Doyle_2019} using new abundance estimates from \cite{Fortin_Archambault_2020}. Solar System objects are shown for comparison on the left.}
\label{bargraph_16}
\end{center}
\end{figure*}

The characterization of rocky exoplanets is a growing area of research. Still, relatively little is known about the compositions of extrasolar rocky bodies, in part because it is the atmospheres of exoplanets that are  amenable to direct observations, and compositions deduced from mass-radius relationships are highly degenerate and have limited precision \citep[e.g.][]{Dorn2015}. A powerful technique for acquiring detailed geochemical information about rocky exoplanets is the analysis of polluted white dwarf (WD) stars. White dwarfs are the last observable stage of stellar evolution for stars where $M_{*}$ $<$ 8$M_\odot$. Because of the extraordinary gravity associated with these electron degenerate stars, elements heavier than helium sink rapidly below their surfaces. Yet, spectroscopic studies show that 25 - 50\% of WDs are `polluted', exhibiting elements heavier than helium \citep{RN16, RN17, RN19}. The source of these heavy elements in WDs is exogenous, the result of accretion of rocky debris from parent bodies that previously orbited the WDs \citep{RN9, Veras2016, Farihi_2016, RN22}. Direct evidence for the pollution being of planetary origin exists in the form of infrared (IR) excesses in the spectral energy distributions (SEDs) of the white dwarfs \citep{Zuckerman_1987, Becklin2005, Jura_2007}. The IR excesses emanate from circumstellar dust disks that thermally reprocess the light from the star \citep{RN7}. It is estimated that $\thicksim$4$\%$ of white dwarfs have detectable IR excess attributed to circumstellar debris disks \citep{RN87, Farihi2009, Barber2012}. An even smaller number of white dwarfs exhibit disks of gaseous material in Keplerian rotation \citep[e.g.][]{Gansicke2006} that has been photoionised by the white dwarf \citep{Melis_2010}. Other direct evidence for the pollution being of planetary or asteroidal origin comes from detection of rapidly evolving transits, although to date only a handful of white dwarfs are observed to have a transiting body \citep{Vanderburg2015, Manser_2019, Vanderbosch2019}.

Spectroscopic studies of polluted WDs allow us to analyze the elemental constituents of extrasolar rocky bodies and evaluate the geochemical and geophysical nature of exoplanets in general. Indeed, a number of authors have applied these principles to the dozen or so currently well-studied polluted WDs where the four major rock-forming elements are detected and quantified \citep{RN10}. It is possible that WDs are accreting multiple small planetary bodies \citep[e.g.][]{RN7}. In these cases, mixing of material would minimize compositional extrema. In this work, we assume pollution occurs from the disruption of single bodies. Generally speaking, the compositions of the bodies polluting WDs resemble those of rocky bodies in our own solar system (Figure \ref{bargraph_16}). While many of the observations have so far been consistent with dry, rocky parent bodies, there are a few important exceptions. One WD exhibits a pattern of abundances with unusually high O, C, and N that is best explained as an extrasolar Kuiper-Belt-analog \citep{RN91}, and more recently, a white dwarf  that is accreting material consistent with a giant planet in a close orbit around the star has been observed \citep{G_nsicke_2019}. Also, a handful of WDs exhibit evidence for accretion of water-rich parent bodies \citep{RN21, Jura2009, RN693, RN4, RN12}. This ability to classify the geochemistry of rocky extrasolar bodies can be taken to even greater specificity  by calculating the oxidation states of the parent bodies \citep{Doyle_2019}.

Oxygen fugacity ($f_{\rm O_2}$) is the effective partial pressure of O$_2$ in a system and a commonly used measure of overall oxidation state for rocks. Oxygen fugacity is as important as pressure and temperature in determining the characteristics of a rocky planet. It determines the minerals comprising the rocky mantle of a planet and the abundance of light elements in its metallic core.  These in turn are important factors in determining whether a body will have a magnetic field and how much water can be stored in the interior \citep{Pearson_2014, Frost_2001, Frost_2008, Wood2006, Buffett2000, RN2}. The mineralogy and water content of the mantle will influence the likelihood for plate tectonics and the nature of volcanism that contributes to the atmosphere \citep[e.g.][]{RN64, Schaefer_2017}.

Oxygen fugacity can always be defined thermodynamically to characterize the oxidation state of a rock by evaluating a thermodynamic expression for a reaction between rock components and ${\rm O}_2$. Oxygen fugacities of planetary materials are commonly expressed relative to that for the Iron-W$\ddot{\text{u}}$stite equilibrium reaction between pure metallic iron (Fe) and pure w$\ddot{\text{u}}$stite (FeO). This so-called Iron-W$\ddot{\text{u}}$stite (IW) reference reaction is

\begin{equation}
  \text{Fe} + \frac{1}{2} \text{O}_2 \rightleftharpoons \text{FeO} .
  \label{feo}
\end{equation}

\noindent Relating the oxygen fugacity of a rock to a reference reaction like Equation \ref{feo} allows one to refer to  oxygen fugacity by difference, $\Delta$IW, such that 

\begin{equation}
  \Delta \text{IW} = \text{log} \left( f_{\rm O_2} \right)_{\rm rock} - \text{log} \left( f_{\rm O_2} \right)_{\rm IW}.
  \label{DIWdifference}
\end{equation}

\noindent Presenting oxygen fugacity in this way allows for the comparison of values independent of temperature and pressure, because the standard-state Gibb's free energy of the reaction that contains the pressure and temperature dependence cancels in Equation \ref{DIWdifference}. Only the equilibrium constant composed of the activities of FeO and Fe is required to specify the relative $f_{\rm O_2}$. Comparison between different rocks is facilitated by setting the uncertain activity coefficients to unity, as is common in planetary science. The activity coefficients are effectively the same for all of the rocks and metals considered here, so applying plausible activity coefficients would shift all of the \DIW values shown here systematically by at most  $\thicksim$ one log unit. Therefore, the initial oxidation state of a rocky body with at least some iron metal at the time of its formation is recorded by the concentration of FeO in the rock ($x_{\rm FeO}^{\rm rock}$) and the concentration of Fe in the metal ($x_{\rm Fe}^{\rm metal}$), yielding

\begin{equation}
\Delta \text{IW} = 2 \text{log} \left( \frac{{x_{\rm FeO}^{\rm rock}}}{{x_{\rm Fe}^{\rm metal}}}  \right).
  \label{DIW}
\end{equation}

\noindent Because Equation \ref{DIW} refers to metal-silicate differentiation, it records the oxygen fugacity at the time the planet or planetesimal was forming (differentiating), and is what we refer to as the ``intrinsic" oxygen fugacity of a body. 

In the Solar System, most rocky bodies formed under oxidizing conditions consistent with \DIW $\geq -2$ \citep[e.g.][]{RN24}. However, the oxygen fugacity of a protoplanetary disk, a hydrogen-rich gas of solar composition, is reduced with \DIW $\thicksim -6$ to $-7$ \citep{RN749}. This dichotomy means that while Mercury and enstatite chondrites formed at oxygen fugacities consistent with those of the protoplanetary disk (with \DIW $\thicksim -3$ to $-7$; e.g. \citealt{RN263}), the majority of rocky bodies in the Solar System require an oxidizing mechanism that increases $f_{\rm O_2}$ by approximately 5 orders of magnitude \citep[e.g.][]{RN80}.

In this work we present the intrinsic oxygen fugacities for 16 rocky parent bodies that are accreting, or did accrete, onto white dwarfs.  We evaluate the robustness of the calculated oxygen fugacities  and determine the conditions under which reduced bodies could be observed. The calculation method for \DIW is described in Section \ref{methods}. The oxygen fugacity results and a description of the modeling efforts to evaluate the robustness of the \DIW values is described in Section \ref{results}. In Section \ref{discussion} we discuss our results and investigate the likelihood of observing a reduced body like Mercury using our method. Section \ref{conclusions} provides a brief summary of the conclusions for this study.

\section{Methods} \label{methods}
\subsection{Selection of White Dwarfs}

In \cite{Doyle_2019} we used stringent requirements for the WDs chosen for study. We required that each star had quantifiable abundances of O, Fe, Mg, Si, Ca and Al in their atmospheres and that each star had a debris disk as confirmed by the presence of an IR excess in the SED. In this work we loosen these constraints to increase the number of polluted WDs available for estimating oxygen fugacities of accreted parent bodies. Here we use stars with quantifiable abundances of  O, Fe, Mg, Si and Ca to calculate $f_{\rm O_2}$. We relax the requirement for  data for the abundance of Al. Where only an upper limit for Al is reported, or where no value is given at all, we use the Al/Ca ratio for CI chondrites to scale the elemental abundance of Al to the Ca abundance in the WD \citep{Lodders_2003}. We scale to Ca because Al and Ca have the same volatility in rock. In each of these instances, we assign the element a measurement uncertainty of 0.301 dex, the value obtained by scaling the measurement by a factor of 2. Aluminum is not as abundant in rock as the other major elements, and varying the assumed abundances by a factor of 2 alters \DIW values by 0.1 dex or less. 

We now also include stars that do not have confirmed debris disks. In the current study, 63\% of the WDs (100\% of the Hydrogen-dominated WDs and 50\% of the Helium-dominated WDs) have confirmed IR excesses indicative of debris disks. Two additional Helium-dominated WDs have unconfirmed IR excesses from WISE data, but WISE has been known to give false positives \citep{Dennihy_2020}. Compared to the total fraction of polluted WDs with debris disks, \citep[$\thicksim$4\%;][]{Barber2012}, our sample of WDs preferentially has material in debris disks that has yet to be accreted. 

\begin{figure*}
\begin{center}
 \includegraphics[width=\textwidth]{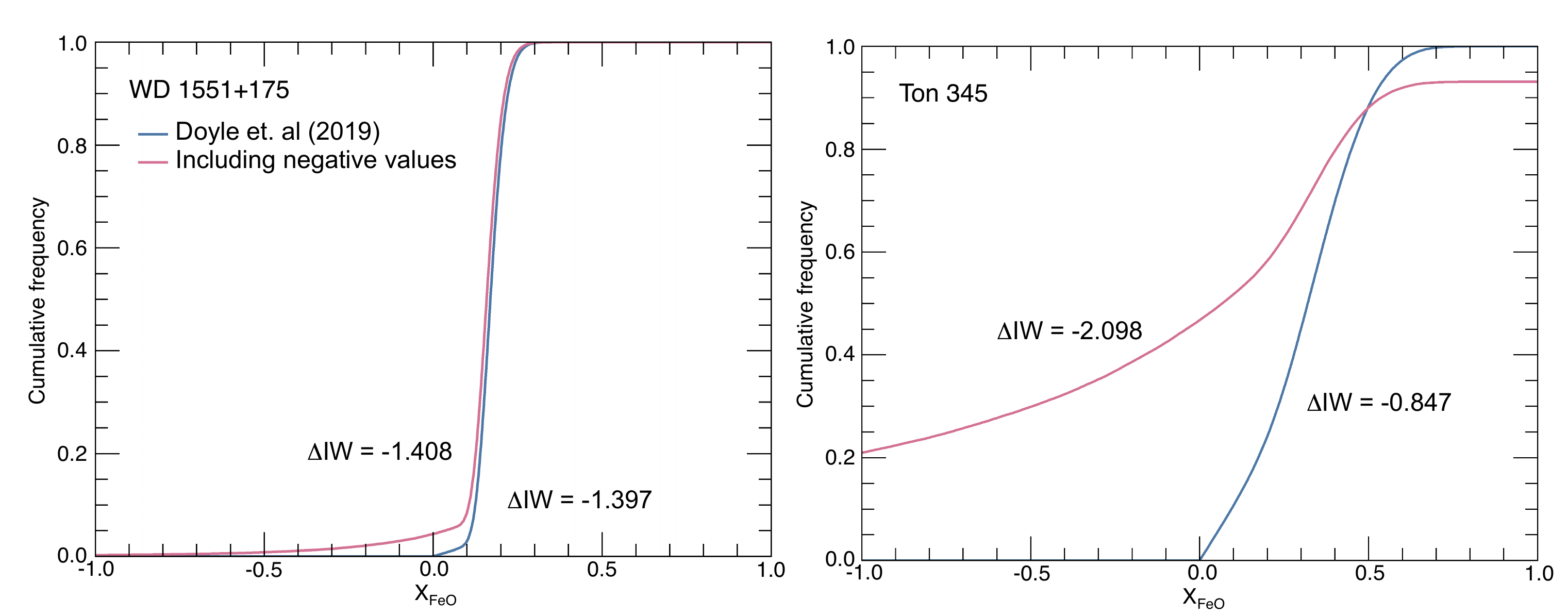}
 \caption{Cumulative frequency distributions for calculated mole fractions of FeO ($x_{\rm FeO}^{\rm rock}$) for two polluted white dwarfs.  On the left are results for WD 1551+175. A well defined $x_{\rm FeO}^{\rm rock}$ is obtained for this case, yielding a well-defined \DIW value. Including negative values for $x_{\rm FeO}^{\rm rock}$ in our Monte Carlo bootstrap method (red) has little effect on the median value for \DIW obtained compared to the value where only positive values are accepted in the Monte Carlo draws for $x_{\rm FeO}^{\rm rock}$ (blue). On the right are results for WD  Ton 345. Including negative values for $x_{\rm FeO}^{\rm rock}$ (red) decreases the median by about 0.25 relative to the value calculated by excluding negative values (blue), resulting in a lowering in the \DIW value of $-1.25$ dex compared with the previous method. These data are sufficient to constrain a median and an upper bound for $\Delta$IW.}
   \label{CDF_neg_pos}
\end{center}
\end{figure*}

\subsection{\texorpdfstring{$f_{\rm O_2}$}\  Calculation}
In order to calculate oxygen fugacities, we follow the methods described by \cite{Doyle_2019}. From the element abundance ratios, we assign oxygen to Mg, Si, Ca, and Al in the necessary proportions to obtain the relative abundances of the charge-balanced rock-forming oxide components MgO, SiO$_2$, CaO, and Al$_2$O$_3$. The remaining excess  oxygen ($\rm O_{\rm xs}$), is assigned to Fe to make FeO until either O or Fe is exhausted.  The excess oxygen available to make FeO is obtained using

\begin{equation}
\frac{\rm O_{\rm xs}}{\rm O_{\rm Total}} = 1 - \frac{\rm O_{\rm SiO_2}}{\rm O_{\rm Total}} - \frac{\rm O_{\rm MgO}}{\rm O_{\rm Total}} - \frac{\rm O_{\rm Al_2O_3}}{\rm O_{\rm Total}} - \frac{\rm O_{\rm CaO}}{\rm O_{\rm Total}},
   \label{Oxs_equation}
\end{equation}

\noindent where ${\rm O_{\it{i}}}$ is the amount of oxygen needed to form the metal oxide, $i$, and ${\rm O_{\rm Total}}$ is the total abundance of oxygen in the system. Other studies have used similar methods for budgeting oxygen \citep{RN10,RN47,RN3,RN4,RN5,RN12}. Once the relative abundances of the oxides are obtained, they are normalized to 1, yielding mole fractions, and permitting application of Equation \ref{DIW}. In principle, if insufficient oxygen exists to pair with Fe to make FeO, the Fe that remains should have been present as metal in the accreted parent body. We emphasize that oxygen fugacity is recorded by the mole fraction of FeO which depends on all of the oxides (FeO, $\rm SiO_2$, MgO, $\rm Al_2O_3$, CaO), and not simply the FeO/Fe ratio for the body. It is possible for metal and water to have coexisted in the parent body if it were undifferentiated, meaning that oxygen which is attributed to FeO in this calculation may have existed as $\rm H_2O$ in the parent body. However, during the differentiation of a rocky body, the oxygen from ices will oxidize metallic Fe to form FeO. We are assuming here that the bodies we are observing in the WDs were either differentiated themselves, or they are the building blocks of differentiated bodies (chondrite meteorites would be the appropriate analog). Where a parent body was composed in part of Fe metal and $\rm H_2O$, our calculation is a measure of the prospects for FeO, and thus the  $\Delta$IW expected for the body taken as a whole, including accreted rock and ices. 

We follow \cite{Doyle_2019} and assign the mole fraction of Fe in the metal core ($x_{\rm Fe}^{\rm metal}$) to  $0.85$ \citep{RN28}. Variations in \DIW of less than 0.2 dex can occur where $x_{\rm Fe}^{\rm metal}$ varies from 0.80 to 1.0, consistent with estimated ranges of bodies in the Solar System and constraints on core-like objects polluting WDs \citep[e.g.][]{Hollands_2018}.

We propagate measurement uncertainties for the polluted WDs using a Monte Carlo bootstrap method with replacement for each elemental abundance. The abundance data are generally reported as log($Z/X$) \citep{Doyle_2019} where $Z$ is the element of interest (O, Fe, Mg, Si, Ca or Al) and $X$ is the dominant atmospheric element for the WD (H or He). We draw single values at random for each element from the lognormal distributions for $Z/X$ as suggested by the symmetrical uncertainties in the log ratio data. A single random draw for each of the elements represents a single instance of the WD data. We calculate a \DIW value from this draw, and repeat the calculation 100,000 times to obtain a frequency distribution of \DIW values. We report median values for \DIW and calculate asymmetrical uncertainties based on an extension of the interquartile range to encompass 67$\%$ of the distribution \citep{Doyle_2019}.

We check our methods by creating fictive polluted WDs in which element ratios for various Solar System rocky bodies are assigned abundances relative to He or H consistent with typical observations (in this case, using GD 40 as the hypothetical WD target). The rocky, Solar System objects used in this study include bulk silicate Earth \citep{RN28}, bulk silicate Mars \citep{RN29}, bulk silicate Mercury \citep{RN31}, Vesta \citep{RN1216}, chondrites (CI \citep{Lodders_2003}, CM, CV, H, L, LL and EH-EL silicate \citep{RN36}). We assign uncertainties to these fictive log($Z/X$) values based on those for GD 40 and propagate measurement uncertainties in the same way as described above. We define the value for $\rm O_{xs}$/Fe for each body to that of GD 40. This value is generally representative of those for most of the WDs in this study (Table \ref{table}). These calculations provide us with a test of the accuracy of the various computational methods described here. The calculated oxygen fugacities for the fictive polluted WDs accreting Solar System objects are comparable to the canonical oxygen fugacities for these objects \citep[e.g.][]{RN263}. 

\subsection{Constraining Upper Limits} \label{upperlimit}

\begin{figure}
\begin{center}
 \includegraphics[width=3.3in]{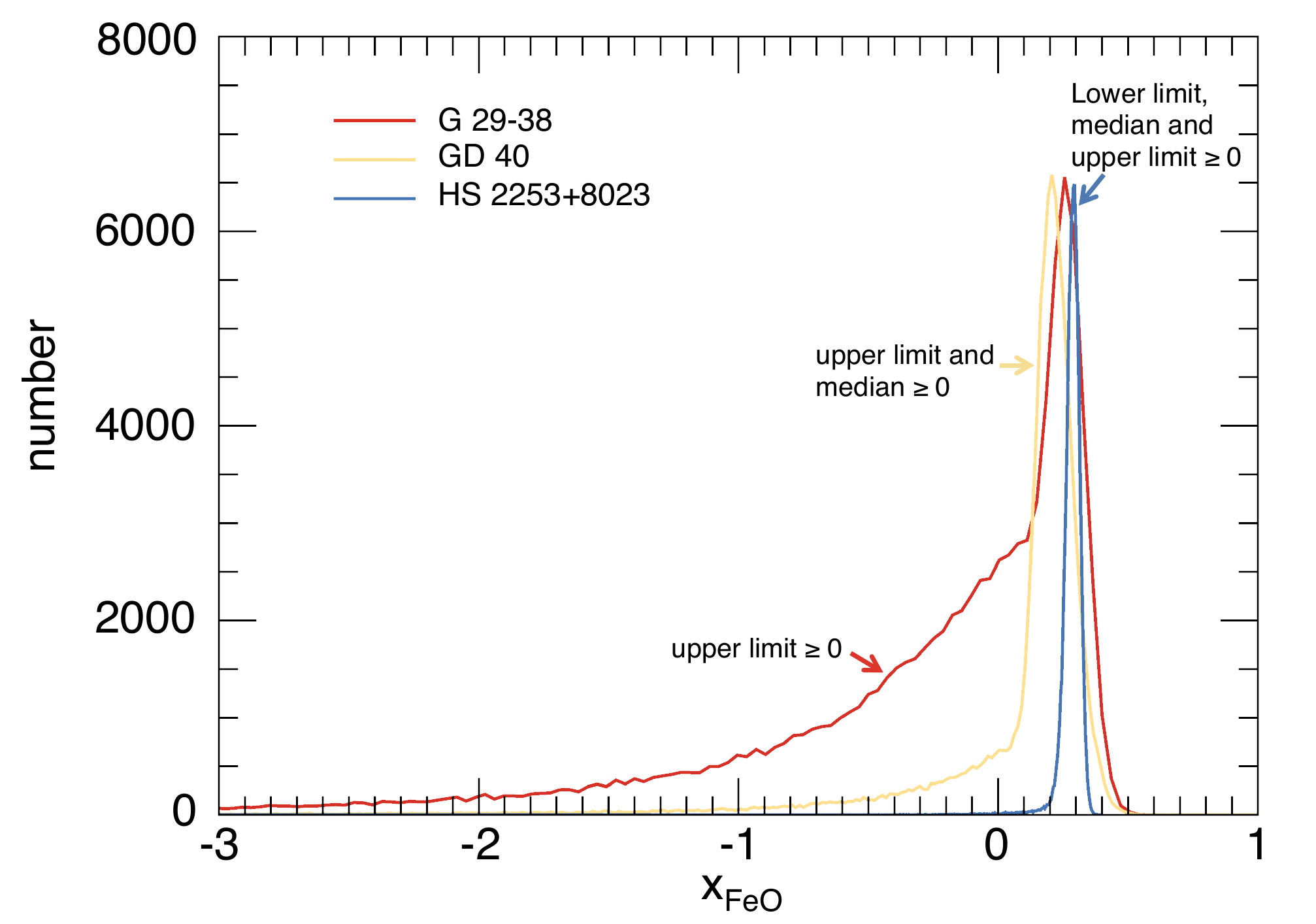}
 \caption{Comparison of parent body FeO mole fractions obtained by Monte Carlo error propagation from WDs that yield an upper limit (red), a median with an upper bound but no lower bound (yellow), and a median with upper and lower bounds (blue) for $x_{\rm FeO}^{\rm rock}$. We assign  upper limits to $x_{\rm FeO}^{\rm rock}$ and the corresponding \DIW values where the Monte Carlo bootstrap method yields a proportion of negative values for $x_{\rm FeO}^{\rm rock}$ $\geq$ 50\%. Where the Monte Carlo bootstrap method yields negative $x_{\rm FeO}^{\rm rock}$ values for between 16.5\%  and 50\% of the draws, we report a median and an upper bound  for $\Delta$IW derived from the $x_{\rm FeO}^{\rm rock}$ values.}
   \label{PDF_upperlim}
\end{center}
\end{figure}

Some of the calculated oxygen fugacities for parent bodies accreting onto WDs are constrained as either medians with upper bounds, or as upper limits in this work. These cases arise when there is a dearth of oxygen and the calculated mole fraction of FeO is small in comparison to the errors \citep{Doyle_2019}. In these cases, the abundance of Fe observed in the WD represents accretion of a significant amount of Fe metal, and errors in all of the elemental abundances accumulate in the calculation of the value for $x_{\rm FeO}^{\rm rock}$ (Equation \ref{Oxs_equation}). These errors propagate to the calculated relative oxygen fugacities (Equation \ref{DIW}).

In \cite{Doyle_2019}, we only included Monte Carlo draws where 0 $\leq$ $x_{\rm FeO}^{\rm rock}$. A negative value for $x_{\rm FeO}^{\rm rock}$ is of course non-physical, but arises where errors are large in comparison to elemental abundances. Negative mole fractions of FeO imply no FeO in the rock component of the parent body because there are more metals than can be accounted for by oxygen (i.e. $\rm O_{xs}$ (Equation \ref{Oxs_equation}) is less than Fe). In these instances, oxygen fugacity is orders of magnitude lower than that of the Iron-W$\ddot{\text{u}}$stite buffer. For the majority of WDs in this study, 11 of 16, random draws infrequently yield negative values for $x_{\rm FeO}^{\rm rock}$ and eliminating these negative values does not significantly affect calculated median $\Delta$IW values. For example, as shown in Figure \ref{CDF_neg_pos}, the median \DIW value for WD 1551+175 changes by  $<$ 0.01 dex if we include draws where negative values for $x_{\rm FeO}^{\rm rock}$ are obtained, as opposed to redrawing those values. 

\begin{figure*}
\begin{center}
 \includegraphics[width=\textwidth]{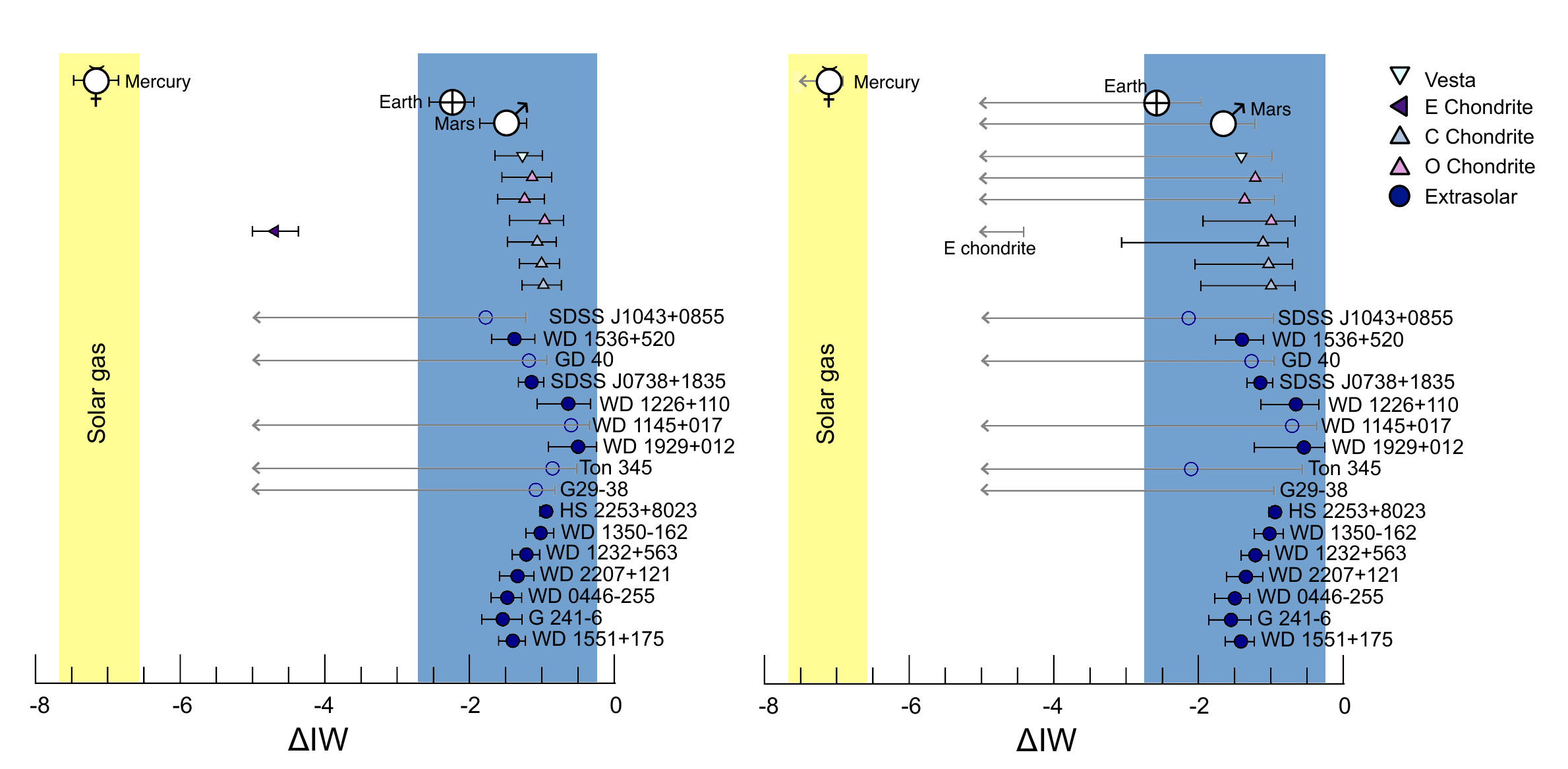}
 \caption{Calculated relative oxygen fugacities for WDs and Solar System bodies in this study.  The left panel shows values obtained using the methods described in \cite{Doyle_2019} and the right panel shows values obtained using the  methods outlined in Section \ref{methods}. Median oxygen fugacities relative to IW are recovered for all but one of the WDs. Solid blue circles are calculated oxygen fugacities for rocky extrasolar bodies accreting onto WDs in this study where both upper and lower bounds are obtained. For four of the sources the median value for $x_{\rm FeO}^{\rm rock}$ is $>$ 0 but the lower bound is not (see Section \ref{upperlimit}). In these cases medians are shown as open circles. The method used to obtain the values shown on the panel on the right protects against biasing \DIW values to higher values and allows us to recover lower medians for SDSS J1043+0855 and Ton 345. Ranges of \DIW for a gas of solar composition (yellow) and for most rocky bodies in the Solar System (blue) are shown for comparison. Hypothetical WDs calculated using rocks from Solar System bodies (see Section \ref{methods}) are also shown for comparison.}
   \label{caltech_paper2_old_Hilkeneg}
\end{center}
\end{figure*}

However, for five WDs in this study, lower bounds on \DIW cannot be obtained because the lower bounds on $x_{\rm FeO}^{\rm rock}$, calculated as the 16.5 percentile in the Monte Carlo distribution, are negative, precluding calculation of a corresponding \DIW value (Equation \ref{DIW}). For these we report median values and an upper bound, or, in some cases where the median mole fraction of FeO is negative, just an upper limit. In the cases where negative $x_{\rm FeO}^{\rm rock}$ values are a significant fraction of the Monte Carlo draws, we improve upon the method of \cite{Doyle_2019} by including $x_{\rm FeO}^{\rm rock}$ $\leq$ 0. Excluding large numbers of negative random draws from our analysis biases the mole fractions of FeO and therefore the median and/or upper bounds on oxygen fugacities derived from them, to spuriously high values. Including the negative $x_{\rm FeO}^{\rm rock}$ in these cases has a pronounced effect on \DIW values. For instance, the calculated median \DIW for Ton 345 changes by $-1.25$ dex when negative values are included (Figure \ref{CDF_neg_pos}).

Here we assign upper limits to WDs where $\geq$ 50\% of the the Monte Carlo bootstrap draws (with replacement) yield  negative values for $x_{\rm FeO}^{\rm rock}$. Where the Monte Carlo bootstrap method yields negative $x_{\rm FeO}^{\rm rock}$ values for between  16.5\%  and 50\% of the draws, we report a median and an upper bound for $\Delta$IW derived from the $x_{\rm FeO}^{\rm rock}$ values (Figure \ref{PDF_upperlim}). Using these criteria, we find one WD for which \DIW values are constrained only by an upper limit (G 29-38) and four WDs for which we report both a median and an upper bound on the \DIW values (SDSS J1043+0855, WD 1145+017, GD 40, and Ton 345). Note that new abundance estimates of rocky debris accreting onto WD 1145+017 from \cite{Fortin_Archambault_2020} yield a median and an upper bound for the derived \DIW values, updated from \cite{Doyle_2019}. The \DIW values for the body accreted by GD 40 is now reported as a median and an upper bound, updated from \cite{Doyle_2019}, because the frequency of negative draws for $x_{\rm FeO}^{\rm rock}$ is very close to the limit we define here (16.5\%).

\section{Results} \label{results}
\subsection{Oxygen Fugacity Values}

\begin{table*}
\caption{\label{table}Oxygen fugacities determined from WD data in this study.}
\begin{center}
{\renewcommand{\arraystretch}{1.5}
\begin{tabular}{l|ccccc}
WD&Type&Temperature (K)&\DIW&$\rm O_{\rm xs}$/Fe&References\\\hline
SDSS J1043+0855&DAZ&18330&$-2.132^{\rm +1.169}$&$0.337^{\rm +10.427}_{\rm -15.209}$&\cite{RN11}\\
WD 1536+520&DBAZ&20800&$-1.374^{\rm +0.284}_{\rm -0.317}$&$6.280^{\rm +6.271}_{\rm -3.705}$&\cite{RN5}\\
GD 40&DBZA&15300&$-1.262^{\rm +0.309}$&$2.495^{\rm +2.629}_{\rm -2.242}$&\cite{RN8}\\
SDSS J0738+1835&DBZ&13950&$-1.137^{\rm +0.167}_{\rm -0.182}$&$10.110^{\rm +8.577}_{\rm -5.269}$&\cite{RN1}\\
WD 1226+110&DAZ&20900&$-0.630^{\rm +0.307}_{\rm -0.432}$&$3.360^{\rm +5.396}_{\rm -2.210}$&\cite{RN6}\\
WD 1145+017&DBZ&14500&$-0.700^{\rm +0.346}$&$1.215^{\rm +4.100}_{\rm -1.650}$&\cite{Fortin_Archambault_2020}\\
WD 1929+012&DAZ&21200&$-0.493^{\rm +0.255}_{\rm -0.411}$&$2.023^{\rm +4.200}_{\rm -1.486}$&\cite{RN6}\\
Ton 345&DBZA&19780&$-2.098^{\rm +1.544}$&$-0.098^{\rm +1.428}_{\rm -1.476}$&\cite{Wilson_2015}\\
G 29-38&DAZ&11820&$< 0.947$&$-0.183^{\rm +1.435}_{\rm -1.413}$&\cite{Xu_2014}\\
HS 2253+8023&DBAZ&14400&$-0.936^{\rm +0.068}_{\rm -0.085}$&$2.858^{\rm +1.151}_{\rm -1.017}$&\cite{RN47}\\
WD 1350-162&DZAB&11640&$-1.012^{\rm +0.182}_{\rm -0.205}$&$4.413^{\rm +2.533}_{\rm -1.878}$&\cite{Swan_2019}\\
WD 1232+563&DZBA&11787&$-1.209^{\rm +0.182}_{\rm -0.194}$&$15.230^{\rm +9.915}_{\rm -6.435}$&\cite{Xu_2019}\\
WD 2207+121&DBZA&14752&$-1.332^{\rm +0.229}_{\rm -0.249}$&$7.271^{\rm +6.574}_{\rm -4.076}$&\cite{Xu_2019}\\
WD 0446-255&DZAB&10120&$-1.473^{\rm +0.198}_{\rm -0.223}$&$4.570^{\rm +3.719}_{\rm -2.603}$&\cite{Swan_2019}\\
G 241-6&DBZ&15300&$-1.536^{\rm +0.266}_{\rm -0.289}$&$7.368^{\rm +5.797}_{\rm -3.791}$&\cite{RN8}\\
WD 1551+175&DBZA&14756&$-1.397^{\rm +0.178}_{\rm -0.195}$&$6.377^{\rm +5.815}_{\rm -3.705}$&\cite{Xu_2019}\\
\end{tabular}
}
\end{center}
\end{table*}

Results show that most rocky bodies accreting onto WDs in this study formed under oxidizing conditions, as did most of the bodies in the Solar System, including Earth (Figure \ref{caltech_paper2_old_Hilkeneg}). Table \ref{table} lists the median values for \DIW and $\thicksim1\sigma$ errors (from an extension of the interquartile range to encompass 67$\%$ of the distribution) for each WD in this study. Variations in \DIW of $\thicksim$ 0.2 dex beyond those portrayed in Table \ref{table} can occur where $x_{\rm Fe}^{\rm metal}$ varies from 0.80 to 1.0. The values reported in Table \ref{table} assume $x_{\rm Fe}^{\rm metal} = 0.85$. In Figure \ref{caltech_paper2_old_Hilkeneg}, results of WDs using methods from \cite{Doyle_2019} are compared to results from this study. WDs with \DIW values represented by dark blue circles show the results where medians and both upper and lower bounds are obtained. Open circles show instances where medians with upper bounds are obtained. In one instance, G 29-38, the median value for $x_{\rm FeO}^{\rm rock}$ is negative when including all random draws, therefore only an upper limit with no median is reported.

\begin{figure*}
\begin{center}
 \includegraphics[width=\textwidth]{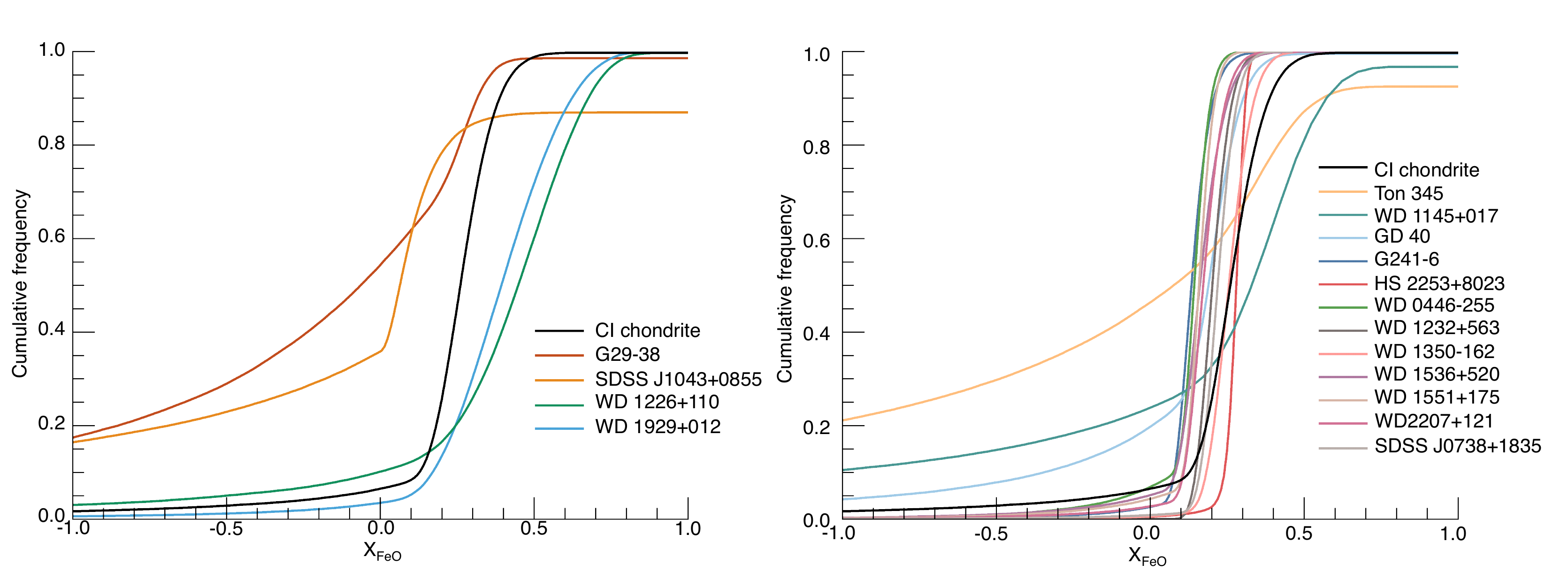} 
 \caption{Cumulative frequency distribution functions for the Hydrogen-dominated (left) and Helium-dominated (right) WDs used in this study. Values for \DIW are constrained as upper limits where the frequency of negative $x_{\rm FeO}^{\rm rock}$ is $\geq$ 50\% of total draws and medians with upper bounds exist for WDs where the frequency of negative draws is such that 16.5\% $\leq$ $x_{\rm FeO}^{\rm rock}$ $\leq$ 50\%. For these five WDs for which lower bounds are not recoverable, we include negative values in the calculation of the $x_{\rm FeO}^{\rm rock}$ distribution in order to avoid biasing the calculated oxygen fugacity. Fictive white dwarfs accreting a debris disk composed of CI chondrite are shown for comparison in each panel (see text).}
   \label{CDF_DAs_DBs}
\end{center}
\end{figure*}

Figure \ref{CDF_DAs_DBs} shows cumulative distribution functions for all of the WDs in this work. One can see the frequencies of negative $x_{\rm FeO}^{\rm rock}$ values for  G 29-38 (DA), SDSS J1043+0855 (DA), GD 40 (DB), Ton 345 (DB) and WD 1145+017 (DB) are greater than for the other WDs. These are the five WDs for which lower bounds on the oxygen fugacities are not constrained.  In four of these five, the frequency of negative draws for $x_{\rm FeO}^{\rm rock}$ is between 16.5\% and 50\%, and a positive median is recovered. In one case, G 29-38, negative mole fractions are obtained in more than 50\% of the draws, and a positive median $x_{\rm FeO}^{\rm rock}$ is not recovered, resulting in only an upper limit on the value for \DIW. Inspection of the cumulative frequency distributions suggests that these five WDs may be accreting reduced bodies more akin to Mercury. The broad distributions underscore the significance of including negative values for $x_{\rm FeO}^{\rm rock}$ in order to avoid biasing the result. These cumulative frequency distributions provide an immediately discernible indication of instances where $x_{\rm FeO}^{\rm rock}$ is evidently low, but where errors are also large.

We point out that Earth is more reduced than many bodies in the Solar System, including chondrites, Vesta and Mars. In order to identify Earth-like extrasolar material, we need better \DIW resolution to distinguish Earth-like and chondritic values. In implementing this new method and utilizing negative values for $x_{\rm FeO}^{\rm rock}$, we find that the rocky body accreting onto Ton 345 is more reduced than the previous method would have implied. SDSS J1043+0855 shows a similar shift to a lower \DIW value. We reiterate that utilizing negative values for $x_{\rm FeO}^{\rm rock}$ removes biases that otherwise force relatively low \DIW values towards artificially high values (see Section \ref{upperlimit}), thereby affording more resolution in \DIW values. The greater resolution enables us to differentiate between an Earth and a chondritic \DIW value, for example (Figure \ref{caltech_paper2_old_Hilkeneg}).

$\rm O_{\rm xs}$/Fe ratios in Table \ref{table} provide a good indication of whether a WD will yield a median \DIW with lower and upper bounds, a median \DIW with an upper bound but no lower bound, or just an upper limit on the $\Delta$IW. The value for this ratio is ideally unity for any rock (Equation \ref{Oxs_equation}). Values statistically greater than 1 suggest another source of oxygen, either as the result of differential gravitational settling or perhaps accretion of a volatile species (e.g., H$_2$O).  Values less than 1 suggest another source of iron (metal). Generally speaking, where errors allow $\rm O_{\rm xs}$/Fe to be $<$ 1, lower bounds are not recoverable. As values for $\rm O_{\rm xs}$/Fe decrease, so does the ability to recover this lower bound. Of the 16 WDs reported in Table \ref{table}, only 2 have negative $\rm O_{\rm xs}$/Fe, and the most negative value is the same WD that provides only an upper limit on $\Delta$IW. 

\subsection{The Effects of Settling on Calculated Oxygen Fugacities} \label{SettlingModel}
We have thus far invoked a correction for differential settling of the elements assuming a steady-state between accretion and diffusive settling for the 4 Hydrogen-dominated WDs (DAs) as in \cite{Doyle_2019}. The assumption of a steady state is generally reasonable in these cases because of the relatively short characteristic timescales for settling in these WDs. Because of the comparatively long characteristic timescales for diffusive settling of the heavy elements in helium-dominated WDs (DBs), they are assumed to be in a build-up phase during which the observed element ratios are attributed to the accreting parent body with no corrections for settling. The veracity of this latter assumption varies from star to star \citep{RN50, Girvenetal2012}.

The robustness of the derived oxygen fugacities under these assumptions can be addressed using the effects of a range of possible scenarios for accretion of material onto a star together with diffusion of heavy elements out of the stellar atmosphere. These two processes can interact to affect the observed elemental abundances in a WD atmosphere. Three different accretion/diffusion phases for planetary debris accreting onto a WD have been described previously \citep{Dupuis_1992,Dupuis_1993,RN50} and include an initial build-up phase of elements into the atmosphere as the debris accretes onto the star, followed by a steady state between accretion onto the WD and settling of elements into the star, and finally a declining phase as the flux of accretion decreases and gravitational settling out of the photosphere of the WD dominates. A combination of mass and ionization state generally leads to heavier elements sinking faster than lighter elements (e.g. iron will sink faster than oxygen) with some exceptions. Over time, these differences in diffusive velocities lead to changes in abundance ratios. The effects of settling on observed element ratios can be accounted for using models of various levels of complexity. 

Extracting element ratios for the accreting parent body from the ratios measured in the WD is most easily done when accretion is clearly either in the build-up phase or the steady-state phase \citep{Dupuis_1992}. In these cases a simple mass balance expression for the mass of element $z$ in the convective layer (CV) in terms of a constant rate of accretion, $\dot{M}_{{\rm CV},z}$, and the rate of gravitational settling based on a constant diffusive velocity can be used \citep{Dupuis_1992}:

\begin{equation}
   \frac{  d{M}_{{\rm CV},z} }  {dt}  =
   \dot{M}_{{\rm CV},z} -\frac{  M_{{\rm CV},z}  }{ \tau_{z} }
   \label{conservation}
\end{equation}

\noindent where $\tau_{z}$ is the constant e-folding time for diffusive settling through the base of the convective layer. The general solution is

\begin{equation}
 \begin{aligned}
   M_{{\rm CV},z}(t) & = M^{o}_{{\rm CV},z}{{e}^{-t/{{\tau }_{z}}}} \\
   & +{{e}^{-t/{{\tau }_{z}}}}\int{{{e}^{t/{{\tau }_{z}}}}}{{\dot{M}}_{{\rm CV},z}}(t)\,dt 
 \end{aligned}
 \label{solution1}
\end{equation}

\noindent where $M^{o}_{{\rm CV}}$ is the initial mass of $z$ in the convective layer that we generally assume is zero. When settling times are long compared with the duration of accretion, $t$, and the rate of accretion is essentially constant, Equation \ref{solution1} reduces to 

\begin{equation}
    M_{{\rm CV},z}=\dot{M}_{{\rm CV},z} t.
    \label{shortaccretion}
\end{equation}

\noindent One expects this case to be most suitable for a DB WD at moderate $T_{\rm eff}$ surrounded by a debris disk since the settling e-folding times in this case can be of order $10^5$ to $10^6$ years.  The ratio of heavy elements observed in the WD is in this case a direct measure of the elemental ratio in the parent body (PB) assuming that the ratio of accretion rates accurately reflects the elemental ratio in the parent body. From Equation \ref{shortaccretion} the relationship between the ratios of two heavy elements $z_1$ and $z_2$ in the WD and in the accreting parent body is therefore

\begin{equation}
    \frac{M_{{\rm CV},z_1}}{M_{{\rm CV},z_2}} = \frac{\dot{M}_{{\rm CV},z_1}}{\dot{M}_{{\rm CV},z_2}} = \frac{M_{{\rm PB},z_1}}{M_{{\rm PB},z_2}}.
    \label{buildup}
\end{equation}

\noindent Conversely, when the duration of accretion is relatively long compared with the e-fold time for settling in the WD, $t >> \tau_z$ and ${e}^{-t/{{\tau }_{z}}} << 1$, and the solution to Equation \ref{solution1} becomes

\begin{equation}
 \begin{aligned}
    M_{{\rm CV},z}(t)=& \dot{M}_{{\rm CV},z} \tau_z (1-{e}^{-t/\tau_z}) \\
    =& \dot{M}_{{\rm CV},z} \tau_z .
 \end{aligned}
\end{equation}

\noindent Under these circumstances, the ratio of the masses of two elements in the WD convective layer is related to that in the accreting parent body by 

\begin{equation}
    \frac{M_{{\rm CV},{z_1}}}{M_{{\rm CV},{z_2}}} = \frac{\dot{M}_{{\rm CV},{z_1}}}{\dot{M}_{{\rm CV},{z_2}}}\frac{\tau_{z_1}}{\tau_{z_2}}  = \frac{M_{{\rm PB},{z_1}}}{M_{{\rm PB},{z_2}}}\frac{\tau_{z_1}}{\tau_{z_2}}
    \label{steadystate_ratio}
\end{equation}

 \noindent where again it is assumed that the accretion rate ratio faithfully reflects the elemental ratio of the accreting parent body, and that the rates of accretion are constant. Equation \ref{steadystate_ratio} is the basis for the often applied steady-state correction to WD element ratios used to obtain parent body ratios: 
 
 \begin{equation}
   \frac{M_{{\rm PB},{z_1}}}{M_{{\rm PB},{z_2}}}= \frac{M_{{\rm CV},{z_1}}}{M_{{\rm CV},{z_2}}}\frac{\tau_{z_2}}{\tau_{z_1}}.
   \label{steadystate}
\end{equation}

\noindent In the case of DA white dwarfs, diffusion timescales ($\tau_z$) are on the order of days to thousands of years and are much shorter than typical estimates for the duration of an episode of accretion represented by a polluted WD ($\tau_{\rm disk}= 10^4-10^6$ yrs) \citep{Girvenetal2012}. In these cases, a steady-state phase should be reached rapidly, soon after accretion begins, and we are almost certainly observing the star post build-up phase. Therefore, for DA white dwarfs, we apply the correction from Equation \ref{steadystate} using calculated values for diffusion timescales. Whether the correction should be applied to DB WDs depends on the surface temperature (i.e., $\tau_z$) and estimates for accretion rate and accretion duration, the latter being all but unknown (see Section \ref{uncertainties}). 

\begin{figure}[b]
\begin{center}
 \includegraphics[width=3.3in]{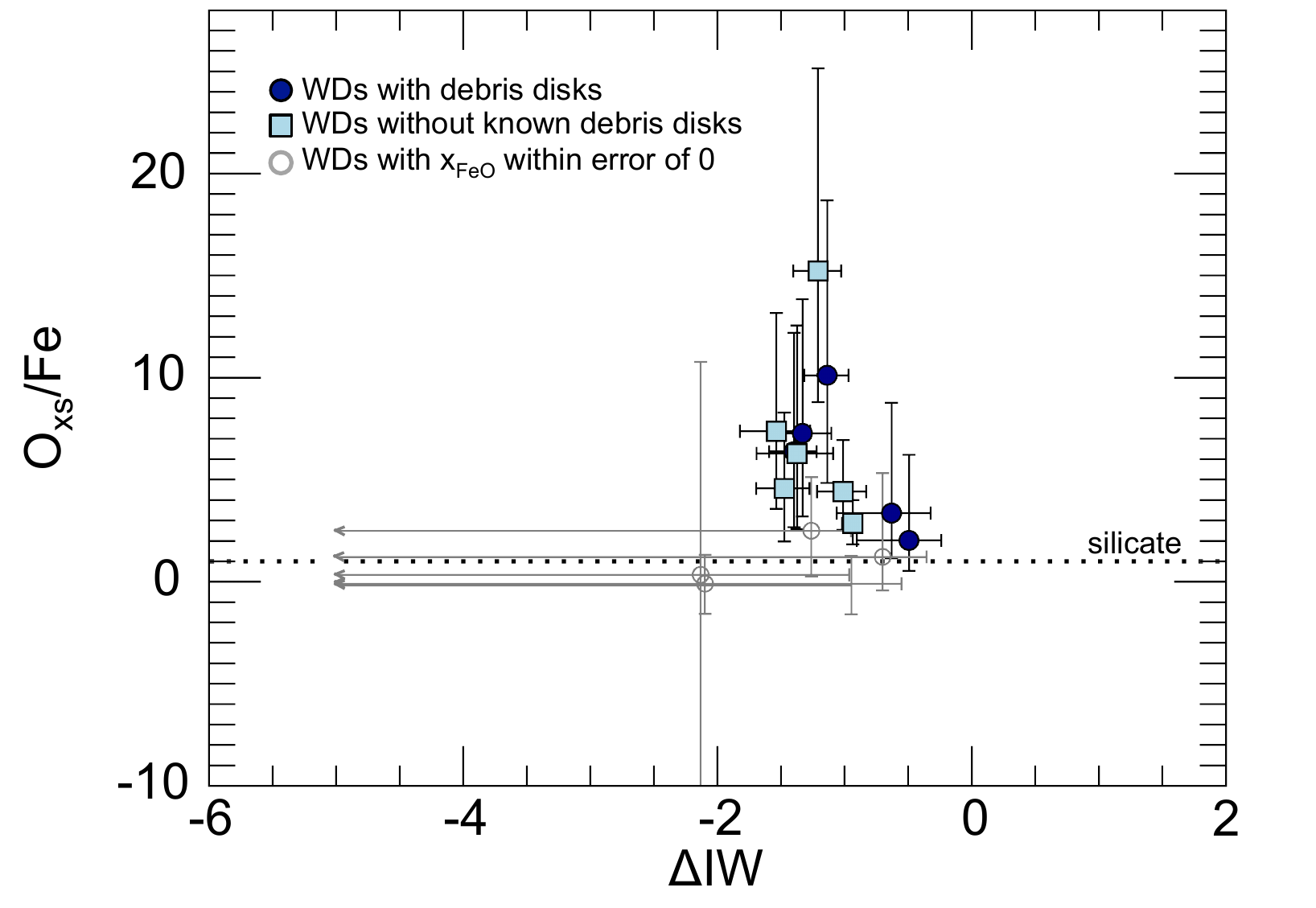} 
 \caption{\DIW vs oxygen in excess of that needed to oxidize Mg, Si, Ca, and Al (O$_{\rm xs}$), relative to Fe. DA and DB white dwarfs with confirmed debris disks are shown as dark blue circles. Light blue squares represent WDs without confirmed debris disks. WDs yielding medians with upper bounds, but no lower bounds for \DIW are shown with open gray symbols. These limits come about when, within errors, there is a dearth of oxygen and FeO is not present in a large fraction of the random draws \citep{Doyle_2019}. The horizontal dotted line represents silicate and oxide rock that, in principle, would always lead to $\rm O_{\rm xs}$/Fe = 1, assuming the amount of Fe$^{3+}$ is small compared with Fe$^{2+}$. WD 1232+563 does not have a confirmed debris disk and has the highest value for $\rm O_{\rm xs}$/Fe in this study, meaning a declining phase of accretion may be a reasonable inference for this WD.}
 \label{Oxs_WDs}
\end{center}
\end{figure}

\begin{figure*}[ht]
\begin{center}
    \includegraphics[width=\textwidth]{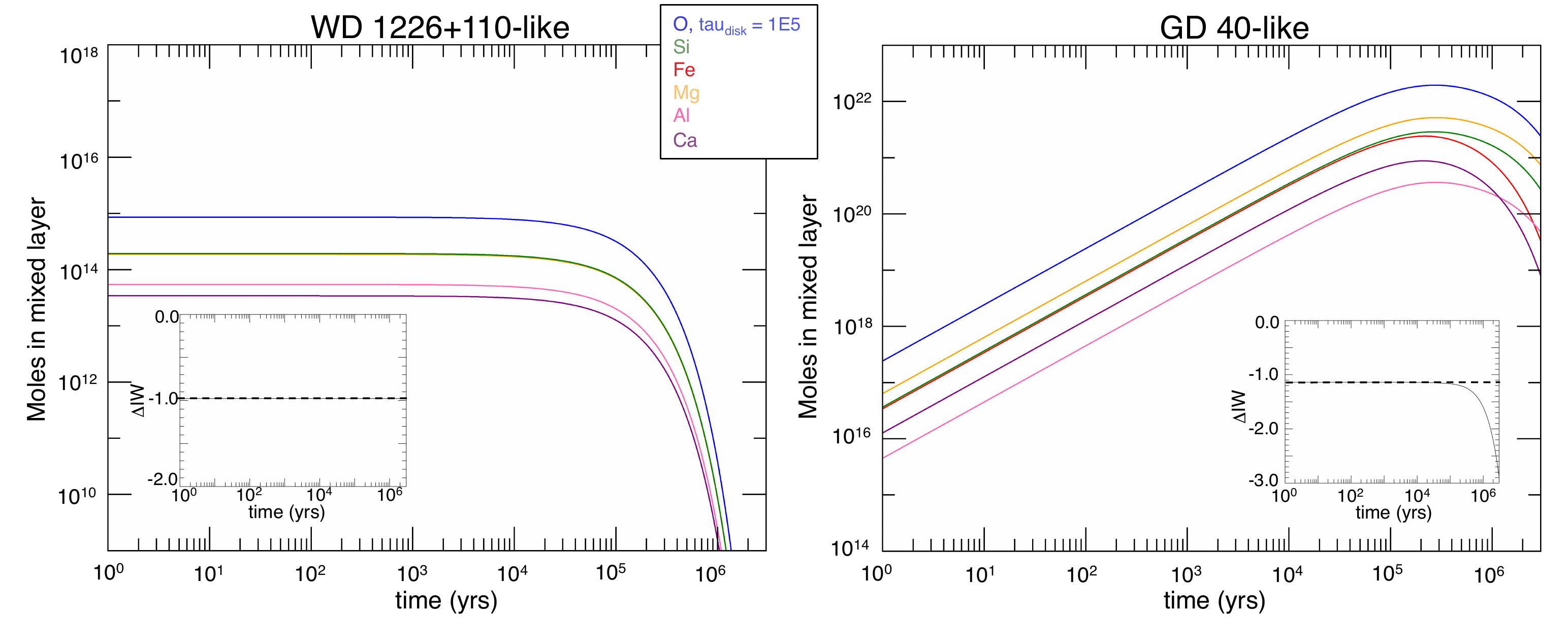} 
    \caption{Model for the time-dependent element accretion history onto a DA similar to WD 1226+110, left, and a DB similar to GD 40, right, using the model for an exponential decay of a debris disk from Equation \ref{Jura_soln}, assuming a debris disk e-folding lifetime of 1 $\times 10^5$ years. Diffusion timescales are from \cite{Koester2013}. The insets show variations in \DIW over this timescale. The parent body \DIW values are indicated as horizontal dashed lines in the insets. The fictive WD 1226+110 initiates with molar compositions corrected using a steady-state approximation (Equation \ref{steadystate}) representative of the parent body. For the DA WD, \DIW is robust throughout all accretion/diffusion phases, because diffusion timescales are short and thus relative diffusion timescales are small. For the DB WD, if the accretion event is longer than $\thicksim$ 2 $\times 10^{6}$ years, one could see a spurious \DIW value indicative of reduced material (\DIW $< -3$) compared with the parent body value of $\thicksim -1$.}
    \label{settling}
\end{center}
\end{figure*}

The inevitable waning of accretion in the declining phase implies that mass accretion rates are not constant (as assumed in Equation \ref{conservation}). As the diffusion of elements out of a WD atmosphere commences and the build-up phase ends, a relative increase in the abundance of oxygen, O, and/or Mg will occur as differences in diffusion velocities cause Si, Al, Ca, and Fe to sink faster. This causes the observed excess oxygen to rise (Equation \ref{Oxs_equation}). Most WDs in this study have oxygen excesses beyond that needed to form FeO. It is important to note that, in these cases, $x_{\rm FeO}^{\rm rock}$ is limited by the abundance of Fe, not oxygen. This is shown in Figure \ref{Oxs_WDs} where the ordinate, $\rm O_{\rm xs}$/Fe,  is $>$ 1. In most cases, element ratios suggest that these excesses are indicative of the accretion of rocky material, including crust (basaltic) and/or mantle without a metallic core. In some cases the oxygen excesses are much larger than can be accounted for by accretion of ices (e.g. $\rm H_2$O, CO) with rock. In these extreme cases, a large value for $\rm O_{\rm xs}$ may be the result of accretion in a post build-up or declining phase. The existence of debris disks around most of the WDs in this study decreases the likelihood that we are observing stars in the declining phase, in which the material in the WD atmosphere is residual, because there is still material in a disk that has yet to be accreted. However, for WDs such as WD 1232+563, which lacks a debris disk and has the highest $\rm O_{\rm xs}$/Fe ratio in Figure \ref{Oxs_WDs}, a declining phase of accretion is a reasonable inference.

Also shown in Figure \ref{Oxs_WDs} is the effect of a paucity of O in a WD in causing an apparent abundance of iron metal, which in turn results in an indeterminate lower bound for \DIW rather than a unique determination. The upper limit arises in the case of apparent metal because in our calculation, $x_{\rm FeO}^{\rm rock}$ is constrained by the smaller of the two abundances, $\rm O_{\rm xs}$ or Fe. Thus, if a parent body has a large bulk Fe content, or a relatively small bulk O content, $x_{\rm FeO}^{\rm rock}$ will be equivalent to $\rm O_{\rm xs}$. From Equation \ref{Oxs_equation} it is clear that the errors associated with excess oxygen are accumulated from those of all of the major rock-forming elements, and are thus considerably greater than those for Fe alone, as described in detail by \cite{Doyle_2019}. These large errors result in the indeterminate lower bounds on \DIW values described above.

The potential importance of  settling emphasizes the need to consider the effects of variable accretion rates on element ratios in the mixing layer of the polluted WDs. \cite{Jura2009} formulated a model for the time-dependent mass of an element $z$ in the convective layer, $M_{{\rm CV},\it{z}}(t)$, in terms of an evolving mass accretion rate. For simplicity, they posited that the accretion rate onto the WD is a function of the mass in a circumstellar accretion disk, and that the e-folding time for the declining mass of the disk attributable to viscous dissipation, $\tau_{\rm disk}$, is a constant. This results in an accretion rate that decreases exponentially with time. The equation that replaces Equation \ref{conservation} under these circumstances is

\begin{equation}
 \frac{dM_{\rm CV,\it{z}}}{dt} = \frac{M^{\rm o}_{\rm PB,\it{z}}e^{\rm -t/\tau_{\rm disk}}}{\tau_{\rm disk}} - \frac{M_{\rm CV,\it{z}}}{\tau_z},
   \label{Juramodel}
\end{equation}

\noindent where $M^{\rm o}_{\rm PB,\it{z}}$ is the initial mass of $z$ in the parent body that forms the circumstellar disk (and the initial mass of $z$ in the disk).  The solution (see Appendix) is

\begin{equation}
   M_{\rm CV,\it{z}}(t) = \frac{M^{\rm o}_{\rm PB,\it{z}}\tau_z}{\tau_{\rm disk}-\tau_z} \left [ e^{- t/\tau_{\rm disk}} - e^{- t/\tau_z} \right ].
   \label{Jura_soln}
\end{equation}

\noindent \cite{RN50} also examined the effects of Equation \ref{Jura_soln} for element ratios in a polluted WD.  Here we use this model to explore the consequences of elemental settling through the WD envelope on calculated oxygen fugacities as a function of time. For the purposes of this work, in the first instance we assume that $\tau_{\rm disk} = 10^5$ yrs. The reasonableness of this assumption is discussed in Section \ref{uncertainties}.

Figure \ref{settling} shows the calculated evolution of moles of elements in the mixing layer of a WD with time, based on Equation \ref{Jura_soln},  using the examples of a DA similar to WD 1226+110 ($T_{\rm eff}$ $\thicksim$ 21000K), and a DB similar to GD 40 ($T_{\rm eff}$ $\thicksim$ 15000K) \citep{RN6,RN8}. In both cases, the parent body oxygen fugacities, expressed as \DIW values, are taken to be near  $-1$, representing relatively oxidized materials in the context of solar system values \citep{Doyle_2019}. For this calculation, we used diffusion timescales from \cite{Koester2013} based on $\log g = 8.0$ cm s$^{-2}$. For the elements accreting onto the model similar to GD 40, one can see the build up of heavy elements in the atmosphere, followed by an interval of steady state (at the maximum of the curves) and finally a period in which the heavy elements settle into the star as accretion from the disk wanes. In the model DA white dwarf (left), short diffusion timescales result in a rapid approach to steady state, as expected, effectively precluding an observation of the build-up phase. In both cases, once accretion from the circumstellar disk begins to decrease significantly, element ratios change markedly, beyond where a simple steady-state correction can compensate. The \DIW values for the model GD 40 decrease rapidly once accretion has decreased enough that settling dominates, causing Fe to decrease in abundance relative to other elements, thereby decreasing $x_{\rm FeO}^{\rm rock}$ and $\Delta$IW. For this star, under these assumed conditions, if the accretion event is longer than $\thicksim$ 2 $\times 10^{6}$ years, the calculated \DIW value would be spuriously low, suggestive of reduced material rather than oxidized material (\DIW $< -3$). On the other hand, the model for the time evolution of \DIW in the model resembling WD 1226+110 reveals that \DIW is an extremely robust parameter. This is because diffusion timescales in DA WDs are short, limiting the relative diffusive separation of the elements. In what follows we concentrate on DB white dwarfs because the oxygen fugacities calculated from these WDs are evidently sensitive to the duration of accretion and diffusion at the time of the observations.

\section{Discussion} \label{discussion}
\subsection{Accretion of an Oxidized Parent Body}

\begin{figure}[ht]
\begin{center}
\includegraphics[width=3.3in]{CI_contour_Koester2013_May29.pdf} 
\caption{Evolution of calculated \DIW values for a model WD accreting a parent body (PB) composed of CI chondrite material (\DIW $= -1$) as a function of both time and effective temperature for the WD. The accretion/diffusion model is that for an exponentially decaying debris disk as in Equation \ref{Jura_soln}, assuming a disk lifetime of $1\times 10^5$ years. Hotter colors are more oxidized \DIW values and cooler colors are less oxidized values. Diffusion timescales are interpolated as a function of $T_{\rm eff}$ from \cite{Koester2013} based on log {\it g} = 8.0 cm $\rm s^{-2}$. As done with helium-dominated WDs in this study, elemental abundance ratios with no steady-state corrections (i.e., Equation \ref{buildup}) were used to calculate \DIW values.}
\label{CI_contour}
\end{center}
\end{figure}

\begin{figure*}
\begin{center}
\includegraphics[width=6in]{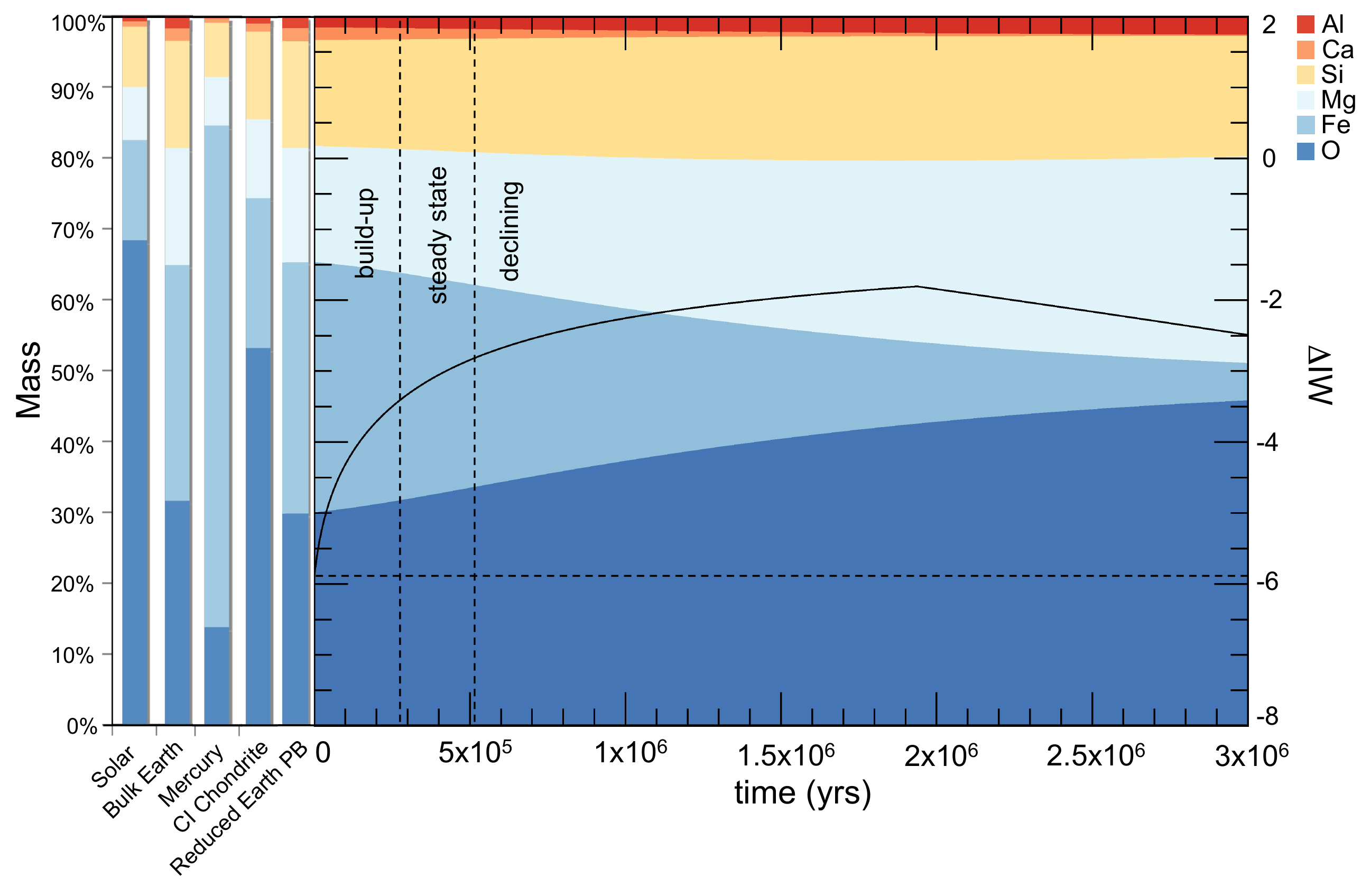} 
\caption{Reduced Earth model depicting variations in abundance ratios with time in the mixing layer of a polluted WD due to differences in diffusion fluxes out of the bottom of the layer at $T_{\rm eff} \sim 15000$K. The time-variable accretion and diffusion model is that defined in Equation \ref{Jura_soln}, and includes a debris disk that decays exponentially with an e-folding time of $1\times 10^5$ yrs. Diffusion timescales are from \cite{Koester2013} and are based on log {\it g} = 8.0 cm $\rm s^{-2}$. The colors represent the normalized abundances of the six major rock-forming elements at each time step in the model. The black line represents the calculated value for \DIW at each time step. The apex of the \DIW curve represents the time at which $x_{\rm FeO}^{\rm rock}$ changes from being constrained by $\rm O_{\rm xs}$ to being constrained by the abundance of Fe. The exact locations of the vertical dotted lines representing shifts in accretion phases are discussed in the text. The reduced Earth model yields \DIW $-5.9$, using our standard method in Section \ref{methods}, and is indicated by the horizontal dashed line. The relative elemental abundances for Solar System bodies and the reduced Earth parent body (PB) are shown for comparison at left.}
\label{barplot}
\end{center}
\end{figure*}

Figure \ref{settling} shows the evolution of material accreting onto two particular WDs. In Figure \ref{CI_contour} we generalize the time-dependent model to consider an exponentially decreasing disk of relatively oxidized CI chondrite material accreting onto helium-dominated white dwarfs varying in temperature from 9000 to 21000K. This range in $T_{\rm eff}$ encompasses all of the He-rich white dwarfs in this study. We interpolate diffusion timescales as a function of $T_{\rm eff}$ again from values calculated by \cite{Koester2013} and based on log {\it g} = 8.0 cm $\rm s^{-2}$. For the purposes of this calculation, we assume the total initial mass, $M_{\rm PB}^{\rm o}$, is equivalent to the mass of asteroid 1 Ceres. For any given temperature, one sees variations in \DIW throughout the various accretion/diffusion phases. For the most part, the model suggests that when oxidized CI chondrites actively accrete onto a DB white dwarf, \DIW should vary with time by less than 0.5 dex from the parent body \DIW value of $\thicksim -1$. However, for DB white dwarfs with effective temperatures of approximately $16000-19000$K, calculated oxygen fugacities for the parent body could decrease to spurious values with \DIW $< -3$ during the declining phase as a result of gravitational settling of elements out of the white dwarf atmospheres. For the assumptions used in these calculations, this phase would begin after about 5$\times 10^5$ yrs, after 5 disk e-folding times. At these $T_{\rm eff}$, if a WD polluted by CI-like material were observed after a million years of accretion with an e-folding time for the debris disk of $10^5$ yrs, the result would be an incorrect conclusion that the parent body rock was relatively reduced rather than oxidized. This is the same phenomenon as seen in the fictive GD 40 in Figure \ref{settling}. In this study, there are no WDs that exhibit such low \DIW values ($< -3$), suggesting that the DBs with $T_{\rm eff}$ in this range are not accreting material in a declining phase millions of years after the start of the accretion event.

\subsection{Accretion of a Reduced Parent Body}

Of the 16 WDs assessed thus far, 11 are consistent with the accretion of oxidized (\DIW $> -3$) rocky bodies, while 5 may be accreting more reduced material. In the Solar System, one terrestrial planet, Mercury, is reduced, as are a small percentage of meteorites (Figure \ref{caltech_paper2_old_Hilkeneg}). If the Solar System were a guide, one might expect to see a Mercury-like oxygen fugacity in $\thicksim 3\%$ of polluted WDs. Here we evaluate the conditions under which the very low $f_{\rm O_2}$ values like those exhibited by Mercury, enstatite chondrites, and aubrites (the igneous equivalent of enstatite chondrites) could be retrieved from a polluted WD using the current method.

Consider the case in which a white dwarf is accreting debris from a disk composed of Mercury, with a large iron core and low FeO in the rock. If we were to assume that the material is accreting onto the star in a build-up phase by equating parent body abundance ratios with observed abundance ratios of the heavy elements, but the material is actually accreting in a steady-state phase, the high relative abundance of O resulting from the unaccounted for settling would cause us to overestimate  $x_{\rm FeO}^{\rm rock}$ and the intrinsic oxygen fugacity for the parent body. This is because there is plenty of iron that is available to pair with the oxygen that has been enhanced by differential gravitational settling, causing a seemingly high FeO concentration by our calculation whereas in reality the iron existed as metal. 

\begin{figure*}
\begin{center}
\includegraphics[width=\textwidth]{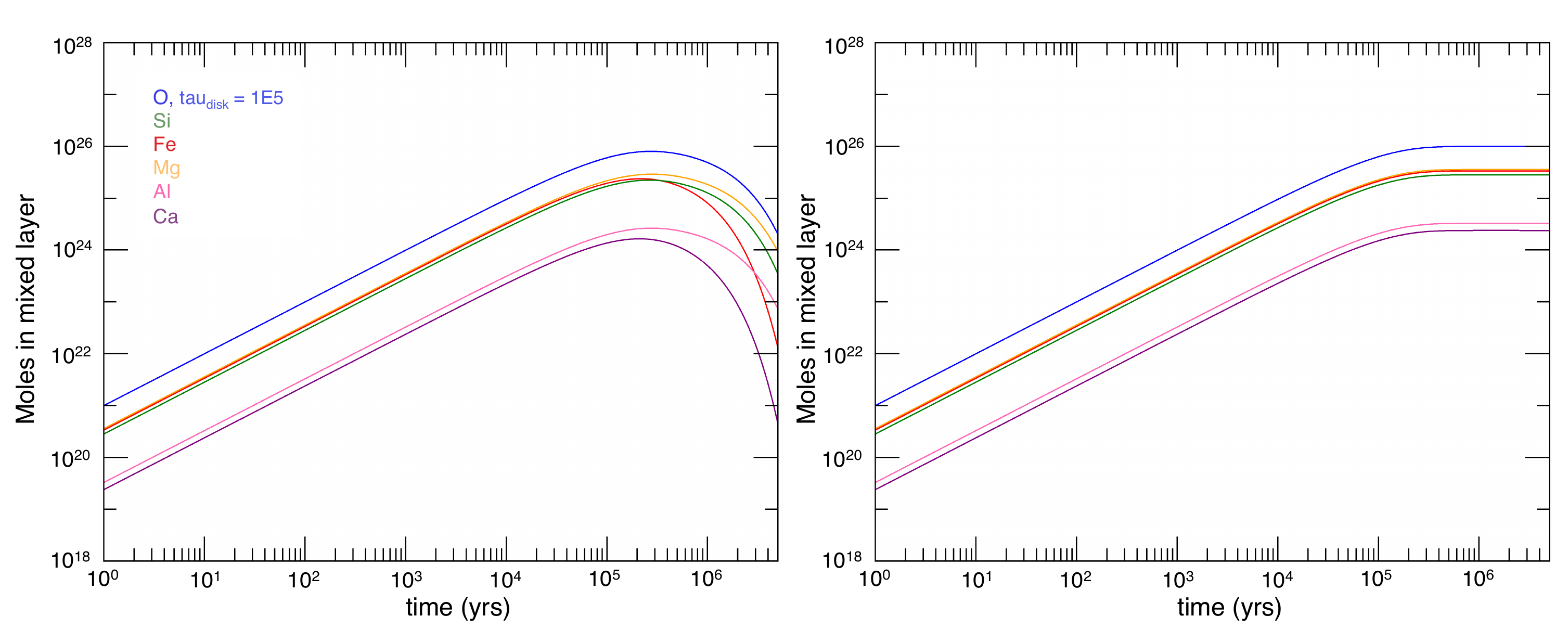} 
\caption{Molar abundances of rock-forming elements in the atmosphere of the same hypothetical WD as that shown in Figure \ref{barplot}. Parent body ratios are the values at time zero. In order to illustrate the build-up phase, we are using log time (unlike Figure \ref{barplot}). The panel on the left represents the time-dependent accretion history of the hypothetical WD as presented in Equation \ref{Jura_soln}. The panel on the right represents the model, but with no gravitational settling through the WD envelope. Note that observable abundances scale with the y-axis. Evaluation of the derivative $dX_O/dt$ in the left panel gives the calculated time at which the build-up phase concludes, $\thicksim$ 3 $\times 10^5$ years. Using the figure on the right, the heavy element concentrations reach their final values  after ~3 to $5 \times 10^5$ yrs, which represents the onset of the declining phase during which the debris disk is depleted.}
\label{nosettle}
\end{center}
\end{figure*}

In order to investigate further  the evolution of \DIW for a reduced body, like Mercury, accreting onto a DB white dwarf, we use a model for a reduced Earth analogous to an enstatite chondrite model for Earth; we do not use Mercury owing to the likely anomalous size of its core \citep{Chau_2018}. Our reduced Earth model has the bulk composition and mass of Earth, but the 8 wt.\ $\%$ FeO in the rock is decreased to $<$1 wt.\ $\%$ by removing the necessary amount of oxygen. The Fe liberated from the oxide is assigned to the metal  core, increasing the core mass from 32 wt.\ $\%$ to 37 wt.\ $\%$. 

Figure \ref{barplot} shows the time-dependent model for an exponentially decaying disk composed of debris from the reduced Earth accreting onto a DB white dwarf where $T_{\rm eff}$ = 15000K. The colors represent the normalized concentrations of the rock-forming elements and the black curve shows the calculated \DIW values where \DIW $= -5.9$ at $t = 0$. One can see that in this  model the reduced body with an abundance of iron metal appears oxidized late in the accretion event.  The observed composition of the body with no accounting for settling seemingly evolves from a mixture of metallic core and rock to a rock-like material anomalously rich in Mg after $10^6$ yrs. We note that this composition influenced by settling does not resemble the compositions in Figure \ref{bargraph_16}.  After 100,000 years into the model, the relative mass of O has increased enough to alter the calculated oxygen fugacity by approximately +2 dex, but the oxygen fugacity would still appear to be relatively low at $\thicksim -4$. The rapid rise in calculated \DIW in the beginning of the accretion episode is because small differences in $x_{\rm FeO}^{\rm rock}$ are more important when the parent body is composed of reduced material where initial values for $x_{\rm FeO}^{\rm rock}$ are small. In the case of the reduced Earth model, initial $x_{\rm FeO}^{\rm rock}$ is $<$ 0.001.

The intervals for the different accretion/diffusion phases in Figure \ref{barplot} are indicated by the dashed, vertical lines. The build-up phase lasts $<$ $3\times 10^5$ years, during which time abundances observed in the WD atmosphere are similar to parent body abundances. The apparent oxygen fugacity increases dramatically even during this stage. We regard the end of the build-up phase as the time at which ${dX_{\rm O}}/{dt}$ = 0 (Figure \ref{nosettle}). We consider the onset of the declining phase to be 5 disk e-folding times, or $5 \times 10^5$ years into the model, which corresponds to the time at which the debris disk is less than 1$\%$ of its initial mass.  The influence of the decrease in mass of the debris disk, and our definition of the declining phase, is clear when the same model as that shown in Figure \ref{barplot} is run with no gravitational settling (Figure \ref{nosettle}). 

The discontinuity at the apex of the \DIW curve occurs at about the initiation of the declining phase at which $x_{\rm FeO}^{\rm rock}$ is no longer constrained by $\rm O_{\rm xs}$ (i.e., where moles Fe $\ge$ moles O) but rather by the abundance of Fe in the WD (i.e. moles Fe $\le$ O; Figure \ref{barplot}). Therefore, if a reduced bulk Earth were to accrete onto this hypothetical DB, its \DIW value would be constrained as an upper limit in our calculation prior to the declining phase, suggestive of the reduced nature of the parent body. However, it would no longer be an upper limit once the declining phase progressed for $1.5\times 10^6$ years. Rather, a moderately oxidized value for \DIW would be obtained until about $3\times 10^6$ yrs of settling, and there would be no signature of the low oxidation state of the parent body during this interval. The likelihood for completely disguising a reduced Earth is high once the pollution episode extends to approximately 5 disk e-folding times or greater in this model, after which point there is no longer evidence that metal existed in the parent body. Similarly, if individual accretion events last for $< 5\times 10^5$ years in the context of this model (i.e., $< 5$ e-folding times for the disk), our calculation method for \DIW should be capable of distinguishing potentially reduced parent bodies from oxidized bodies, even if only by the mere existence of an upper limit constraint and a high propensity for negative mole fractions of FeO.  The uncertainty of disk lifetimes is discussed in Section \ref{uncertainties}.

Generalizing this result, when the disk lifetime is comparable to typical settling times through the mixed layer of the polluted WD, and observations are taken after approximately 5 disk e-folding times or more have passed, calculated oxygen fugacities will be biased against finding reduced parent bodies. In this study, the existence of observable debris disks around most stars decreases the likelihood that we are observing the WDs after several disk e-fold times unless the disks were considerably more massive than is customarily inferred from masses in the mixing layer and estimates of accretion rates (see Section \ref{uncertainties}). Thus, the probability of masking a reduced body is commensurately low.

\begin{figure}[ht]
\begin{center}
\includegraphics[width=3.3in]{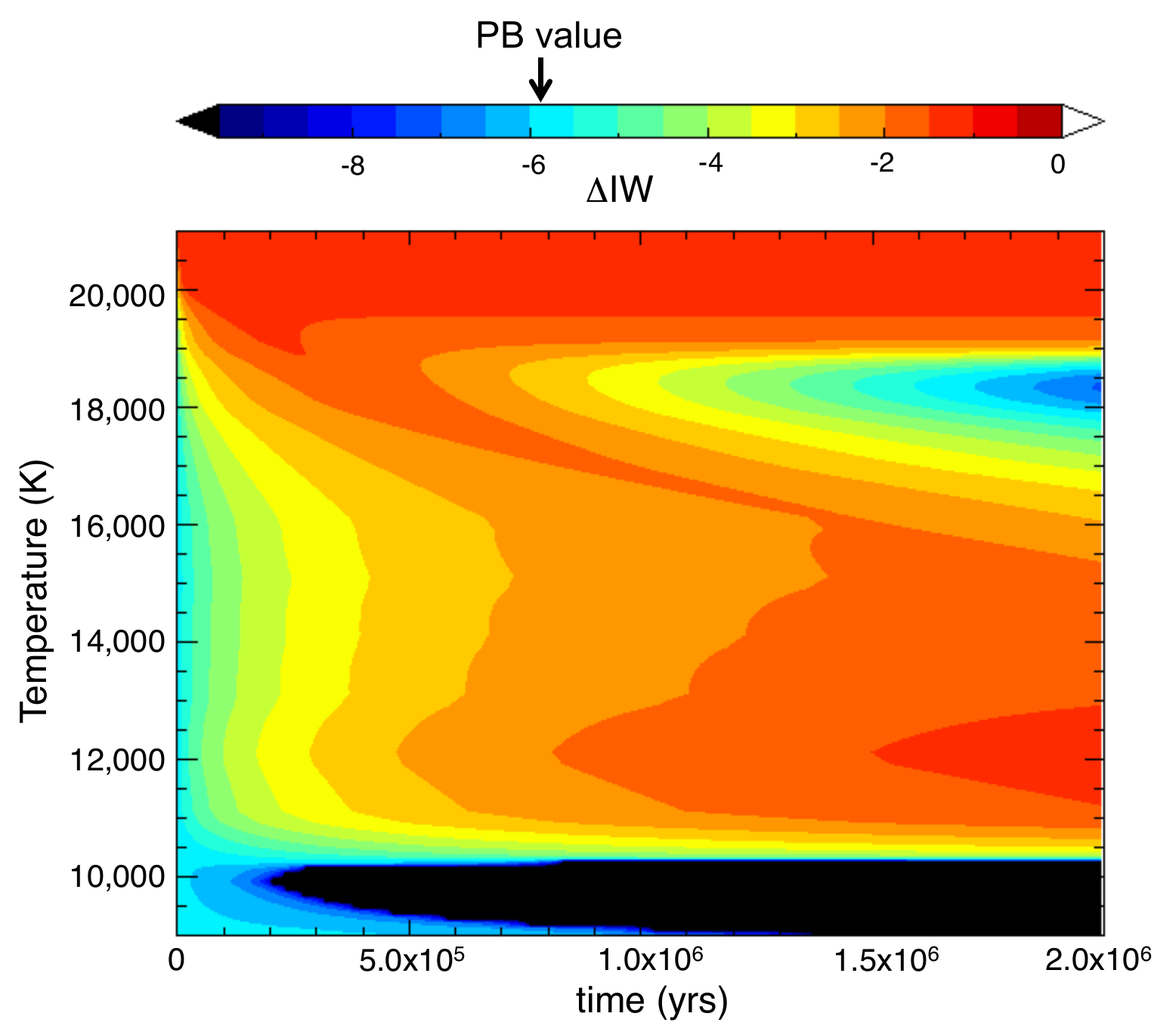} 
\caption{Evolution of calculated \DIW values for  model WDs accreting reduced Earth-like material (see text) as a function of time and $T_{\rm eff}$. The accretion/diffusion model is that for an exponentially decaying debris disk expressed in Equation \ref{Jura_soln}, assuming a disk e-folding lifetime of $1\times 10^5$ years. It is the same model as shown in Figure \ref{barplot}, but expanded to include a range of WD temperatures onto which the parent body accretes. The \DIW value for the model parent body (PB) is $-5.9$, as indicated in the legend. Hotter colors are more oxidized calculated \DIW values and cooler colors are less oxidized. Diffusion timescales are interpolated as a function of $T_{\rm eff}$ from \cite{Koester2013} and are based on log {\it g} = 8.0 cm $\rm s^{-2}$. As done with helium-dominated WDs in this study, elemental abundance ratios with no steady-state corrections (i.e., Equation \ref{buildup}) were used to calculate \DIW values.}
\label{contour}
\end{center}
\end{figure}

In Figure \ref{contour} we show \DIW values obtained by expanding the model for accretion of a reduced Earth shown in Figure \ref{barplot} to include stars varying in $T_{\rm eff}$ from $9000 - 21000$K. As in Figure \ref{CI_contour}, we interpolate derived diffusion timescales for each $T_{\rm eff}$ in DBs from \cite{Koester2013} and invoke a characteristic disk lifetime of $1\times 10^5$ yrs. The variations in \DIW values at T$\thicksim$15000K are equivalent to the black curve in Figure \ref{barplot}.

Figure \ref{contour} shows that low \DIW values are for the most part recoverable for DB WDs with $T_{\rm eff} < 19000$K less than $1\times 10^6$ years into the accretion event (i.e., for ten disk e-folding times). In DBs hotter than approximately 19000K, diffusion timescales become so fast that DB white dwarfs begin to resemble DA white dwarfs in that after an initial episode of oxygen enrichment they rapidly approach a steady state. This is the worst case scenario for detecting low oxygen fugacity parent bodies; for a DB WD with $T_{\rm eff} > 19000$K, preserving low values for \DIW is not possible beyond the steady-state phase. This is true even when using Equation \ref{steadystate}, as is routinely done for DAs, because once the system leaves the steady-state phase, there is no {\it a priori} way to correct for settling combined with waning accretion. This occurs at times greater than about $5\times 10^5$ yrs in Figure \ref{contour}. Below $\thicksim 10000$K, diffusion timescales are longer, closer to the assumed disk lifetime. Therefore, the calculated value for \DIW remains low for the duration of the accretion event well beyond the practical lifespan of the disk (Figure \ref{contour}). Slow diffusion of elements through the white dwarf atmosphere means the relative flotation of oxygen will never dominate over accretion in these cases. In this case, elements do not begin to settle out of sight until the majority of the parent body has been accreted, and the apparent oxygen fugacity decreases because iron is the monitor of $\Delta$IW.

Based on our models, the two DBs with $T_{\rm eff}$ $>$ 19000K in this study (Ton 345 and WD 1536+520) would always appear oxidized if they were accreting in a declining disk phase in which a reduced Earth-like body accretes with $\tau_{\rm disk}$ similar to settling times. However, the data indicate instead that these parent bodies may be reduced rather than oxidized. Ton 345 yields a median parent body \DIW of $-2.1$. This value is lower than could be preserved due to settling unless we were observing the accretion in the first 10$^3$ years of the event. Ton 345 does possess a gas emission disk, which may suggest an early stage of accretion, as these disks may come from colliding fragments of planetesimals \citep{Manser2020}. For WD 1536+520, if one were to calculate \DIW using a steady-state assumption, as considered likely by \cite{RN5}, one gets a median with an upper bound for $\Delta$IW, contrary to the unique determination reported in Table \ref{table} (see discussion surrounding Figure \ref{Caltech_compare}). Here again, evidence for low $x_{\rm FeO}^{\rm rock}$ values at high $T_{\rm eff}$ suggests a short duration of accretion.

\begin{figure}
\begin{center}
\includegraphics[width=3.3in]{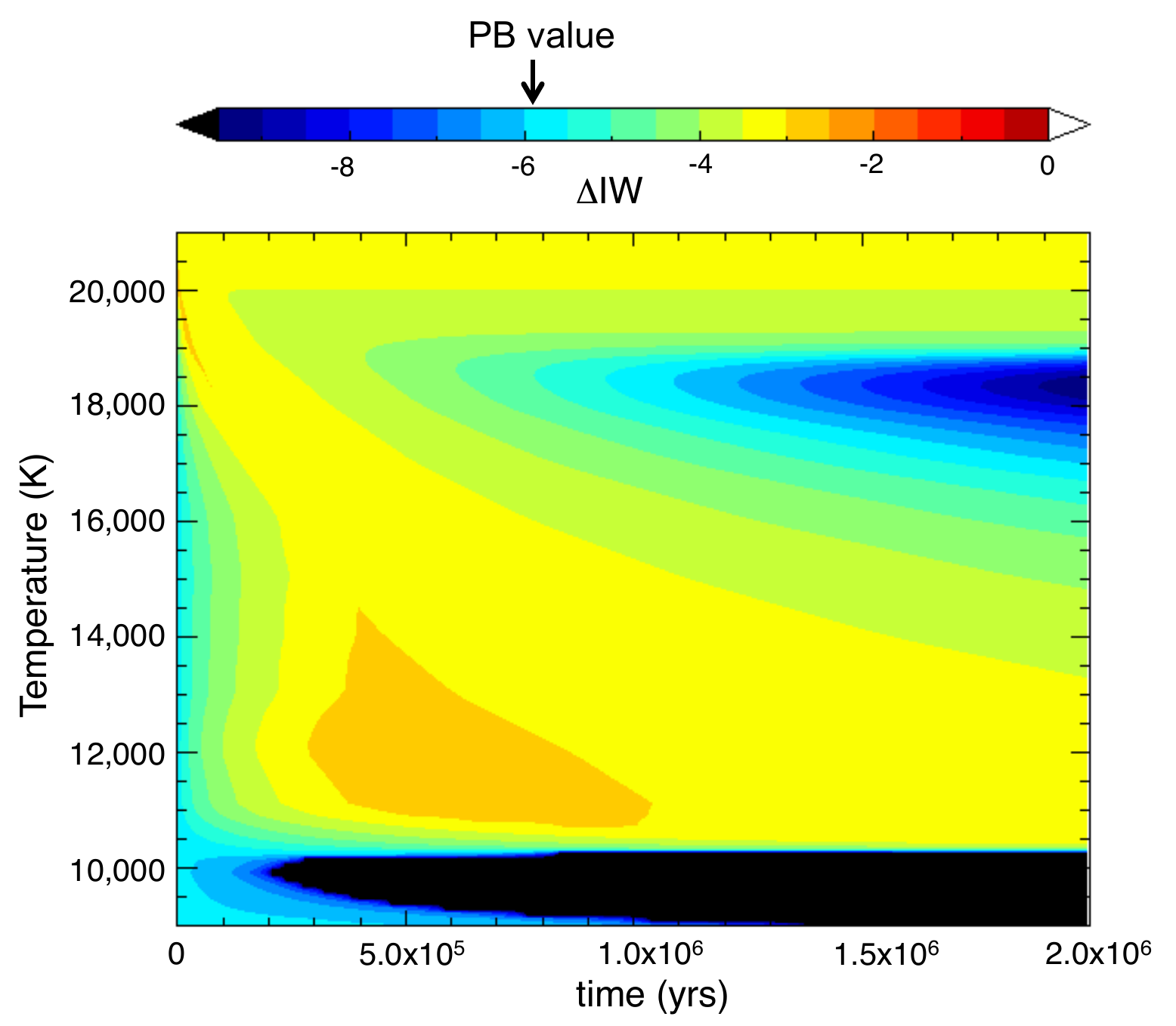}
\caption{Evolution of calculated \DIW values as a function of time and $T_{\rm eff}$ for  model WDs accreting reduced Earth-like material like that shown in Figure \ref{contour}, but with a smaller core mass of 3.7 wt.\ \%. Diffusion timescales are interpolated as a function of $T_{\rm eff}$ from \cite{Koester2013} and are based on log {\it g} = 8.0 cm $\rm s^{-2}$. The \DIW value for the model parent body (PB) is $-5.9$ and the most oxidized value calculated for these fictive WDs is \DIW $-2.86$. With such a small amount of iron metal, there is not enough to pair with floating oxygen to significantly alter $x_{\rm FeO}^{\rm rock}$ and mask the reduced nature of the parent body. As done with helium-dominated WDs in this study, elemental abundance ratios with no steady-state corrections (i.e., Equation \ref{buildup}) were used to calculate \DIW values.}
\label{metal}
\end{center}
\end{figure}

Therefore, when iron metal is present and available to ``pair" with floating $\rm O_{\rm xs}$ using our calculation method in a scenario like that modeled here, it is possible for a reduced parent body to appear as  oxidized in the WD solely due to differences in diffusion timescales. For our model reduced Earth  with a \DIW value of $-5.9$, we find that in order for the calculated \DIW value to signify a reduced parent body with \DIW $\thicksim -3$ or less, irrespective of time of the observation and at any temperature, the object would need to be composed of $<$ 4 wt.\ \% iron metal (Figure \ref{metal}). 

Using these models as guides, one concludes that although identification of oxidized bodies is straightforward, identification of reduced bodies requires a more nuanced approach. For $T_{\rm eff} < 18000$K we should expect to see evidence of low oxidation states in the data for even long accretion times relative to the lifespan of the debris disk. Hints that the WD is in the waning stages of accretion may show up as odd elemental abundances compared with expectations for most rocks, including those evidenced in the polluted WDs (e.g., Mg-rich compositions). Nonetheless, large amounts of metal will always result in large errors for calculated \DIW values. Fortunately, many polluted WDs show evidence of accretion of mainly rock, perhaps because rock and metal are separated during the accretion process \citep[e.g.][]{Manser_2019}. The WDs yielding upper limits in \DIW as the result of an apparent abundance of iron metal are at present the most likely candidates for reduced material (low \DIW values) and should be the focus of future investigation.

\subsection{Uncertainties in the Calculations} \label{uncertainties}

The model shown in Figure \ref{contour} makes clear that crucial parameters necessary to evaluate the veracity of calculated \DIW values for DB WDs are the duration of accretion prior to the observations and the characteristic timescale for depletion of the debris disk. These can not be estimated independently at present and are constrained only with certain assumptions. One can calculate minimum disk lifetimes using the measured masses of all of the rock-forming elements ($Z$, total) in the helium-dominated WD convective layer ($M_{\rm CV, {\it Z},total}$) with calculated accretion rates ($\dot{M}_{\rm CV, {\it Z},total}$). Pairing typical steady-state accretion rates with typical lower limits for parent body masses as evidenced by the masses of heavy elements in the convective layers allows for an estimate of minimum disk lifetimes based on

\begin{equation}
      \tau_{\rm disk} = \frac{M_{\rm CV, {\it Z},total}}{\dot{M}_{\rm CV, {\it Z}, total}}.
      \label{tau_disk_minimum}
\end{equation}

\noindent Using a range for $\dot{M}_{\rm CV, {\it Z}, total}$ of $10^8$ to $10^{11}$ g $\rm s^{-1}$ based on theory and observations \citep[e.g.][]{Rafikov_2011a,Rafikov_2011b,RN30,Wyatt_2014,Xu_2019} and the average masses of rock-forming elements observed in the convective zones  of the DBs in this study one obtains minimum disk lifetimes of $4.5\times 10^4$ to $4.5\times 10^7$ years. Using estimates for the mass of rock-forming elements in polluted DBs from the literature, the same calculation yields minimum disk lifetimes ranging from $3\times 10^4$ to  $5\times 10^6$ years \citep{Girvenetal2012}. These ranges include the nominal value of $1\times 10^5$ years that we used in our modeling.  We emphasize, however, that these estimates rely on the notion that accretion rates are constant. Episodic accretion is not captured by this calculation.

The contours in Figures \ref{contour} will vary as a function of the values for $\tau_{\rm disk}$. The diffusion times, $\tau_z$, for the subset of DBs in this work range from $10^3$ to $10^6$ years. Longer disk lifetimes where $\tau_{\rm disk}$ $>>$ $\tau_z$ lead to more regions in $T_{\rm eff}$ vs.\ time space where the oxidation states of highly reduced bodies would be overestimated. Shorter disk lifetimes where $\tau_{\rm disk} << \tau_z$ lead to greater fidelity in recording the oxidation states of reduced bodies. Our assumed disk characteristic lifetime of $1\times 10^5$ years is a plausible median value, but it is unlikely to be universally applicable. Higher accretion rates, by several orders of magnitude, are obtained when considering the effects of thermohaline mixing and convective overshoot \citep[e.g.][]{Deal_2013,Bauer_2018,Bauer_2019,Tremblay2017,Kupka2018,Cunningham2019}. Higher accretion rates lead to shorter minimum disk lifetimes for the same minimum masses (Equation \ref{tau_disk_minimum}).

Large ranges in accretion rates may imply that accretion is indeed episodic, as has been suggested by others \citep[e.g.][]{RN30, RN12,RN694}. Additionally, the assumption that the accreting material is well-mixed, and therefore representative of the bulk composition of the parent body may be too simplistic \citep[e.g.][]{Xu_2019}. In the future, accretion rates and disk lifetimes will need to be better determined for individual WDs in order to better quantify the reliability of geochemical parameters deduced from polluted WDs.  

\begin{figure}[ht]
\begin{center}
\includegraphics[width=3.3in]{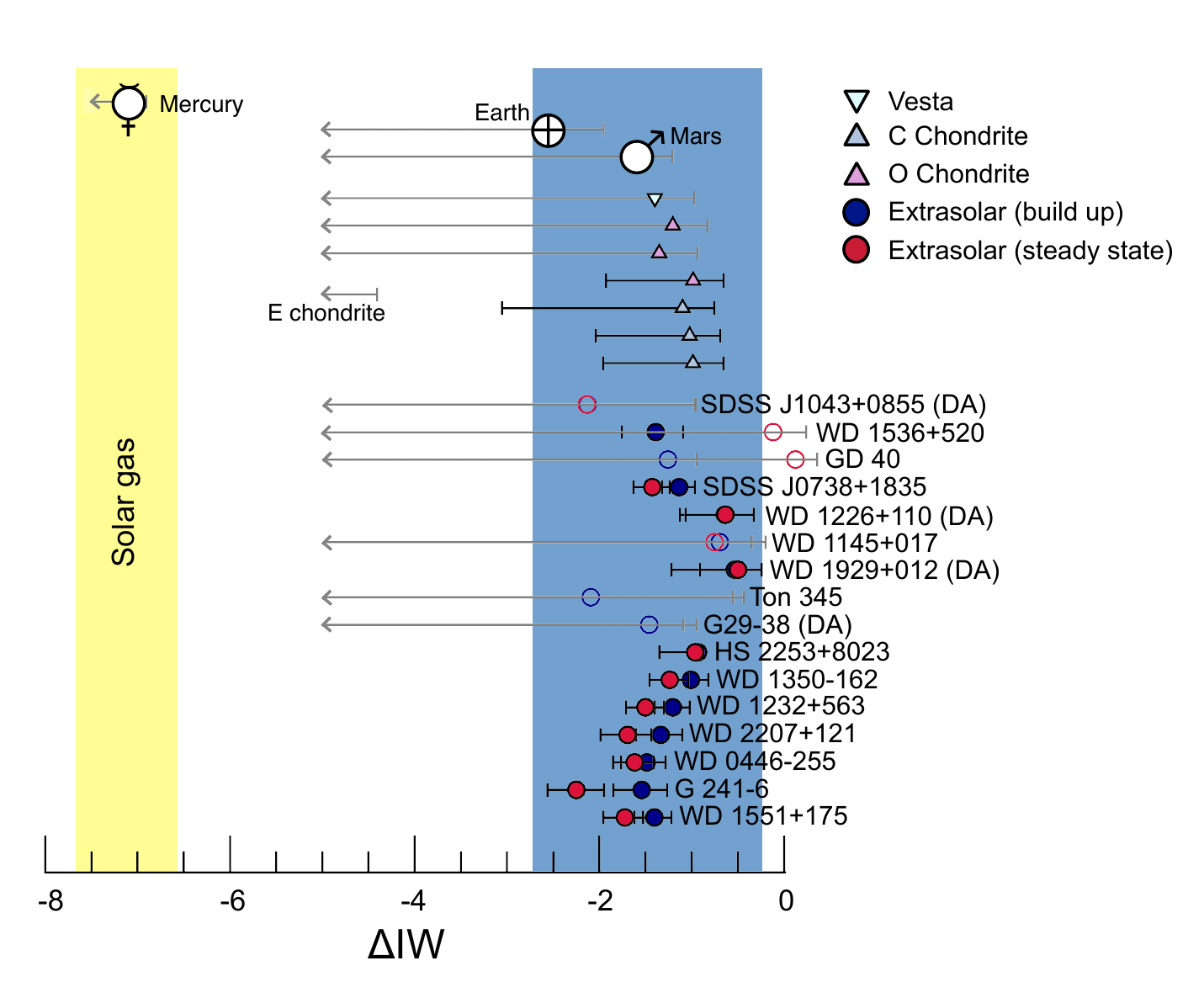} 
\caption{Comparison of calculated \DIW values  assuming that the DBs in this study are accreting in the build-up phase vs. values obtained by assuming they are in steady state between accretion and gravitational settling. Values assuming a build-up phase are represented by unfilled or filled navy circles, depending on whether \DIW is a median with an upper bound (unfilled) or a median with both lower and upper bounds (filled). Values for \DIW for WDs assumed to be in a steady-state phase of accretion are represented in the same way, but with red circles. The ability to calculate the lower bound for WD 1536+520 is lost when steady-state is assumed. DA white dwarfs are indicated by their labels and are generally considered to be in a steady-state phase of accretion. Here we show G 29-38 assuming both a steady-state and a build-up phase of accretion, in consideration of the possibility of thermohaline instabilities in this DA WD (see text). G 29-38 provides an upper limit for \DIW under a steady-state assumption, but we can define the median under the assumption of a build-up phase. The differences in $\Delta$IW between steady-state and build-up assumptions for the other DAs are given in the text. Generally, differences in \DIW when WDs are considered to be in steady state vs. a build-up phase are less than 0.5 dex.}
\label{Caltech_compare}
\end{center}
\end{figure}

Our ability to detect low \DIW values depends on our ability to assess whether a WD is in a build-up, steady-state, or declining phase of accretion.  These assessments can be highly uncertain. Thus, Figure \ref{Caltech_compare} shows the calculated values for \DIW considering both scenarios for DBs, a build-up and a steady-state phase. Generally speaking, for the DB WDs, the median values for \DIW change by $<$0.5 dex when DBs are assumed to be accreting rocky material in a steady state, rather than in the build-up phase that we have assumed previously (Figure \ref{Caltech_compare}). However, in one case (WD 1536+520) assuming a steady-state phase results in the loss of the lower bound on the median \DIW value of $-0.113$ compared to the value in Table \ref{table} of $-1.374$, this is the largest shift in median \DIW values calculated.

For the DA white dwarfs for which element ratios are usually corrected assuming a steady-state phase of accretion, modeling suggests that some of them are candidates for thermohaline instabilities \citep{Deal_2013,Bauer_2018,Bauer_2019}. If thermohaline instabilities are present, measured element ratios should not be corrected with the short settling times typically used. As an illustration, in Figure \ref{Caltech_compare} we show results for DA G 29-38 with and without settling time corrections. Element ratios in G 29-38 do not define a median \DIW value when considered to be in steady state, but do define a median when a build-up phase is assumed (i.e., no corrections are applied). Under the assumption of a build-up phase due to thermohaline instabilities, two of the DAs in this study (WD 1226+110 and WD 1929+012) produce \DIW medians of $-$1.616 and $-$1.948, respectively, and upper bounds of  $-$0.696 and $-$0.673, respectively. The median \DIW for SDSS J1043+0855 is lost when a build-up phase is assumed, but the upper limit decreases to \DIW $= -$2.089. Given the plenitude of WDs in this study with debris disks, accretion of material in a declining phase is less likely, with the possible exception of WD 1232+563, as discussed in Section \ref{SettlingModel}.

\subsection{Implications for Oxidation States in Planetary Science}

The process that gave rise to the oxidation of most rocky bodies in the Solar System, relative to a hydrogen-rich gas of solar composition, is unresolved. It is possible that different processes are involved at different stages of planet formation. During rock formation, the presence of water could increase $f_{\rm O_2}$ in dust that eventually forms planetesimals. After rock formation but prior to planetary assembly, aqueous alteration in the parent body could further increase $f_{\rm O_2}$ as iron metal is oxidized to form iron oxides and iron-bearing clay minerals. Finally, the addition of Si and/or O to the metal core of a planetary embryo or planet would result in the partitioning of Fe into the rock as FeO, and would also lead to increases in the initial intrinsic oxidation state of the rock. These multiple paths to oxidation have been well-studied through modeling efforts, though it is not known which process is dominantly responsible for oxidizing bodies in the Solar System \citep{RN749,RN24,RN80,RN1155}.

Most of the rocky bodies accreting onto the WDs in this study are consistent with oxidizing conditions at the time and place of parent body formation. This could imply that there is something about rock formation in general that may favor oxidation relative to a solar gas and that the formation of planets with Earth-like oxygen fugacities may be more common than forming planets with Mercury-like oxygen fugacities.

However, the difficulty in detecting and accurately measuring the oxidation states of reduced bodies could mean that more reduced bodies exist and we have not yet detected them. A relationship between oxidation state of rock and distance from the Sun has been suggested for the Solar System \citep{Wade2005,Wood2006,Rubie2011,RN1155,Badro2015}. If this relationship where reduced bodies form  preferentially closer to the Sun is typical, then it could be that reduced bodies are destroyed by their host star when the star is on the red giant branch \citep[e.g.][]{Shroder_2008}. This would further decrease the likelihood of discovering a reduced body by the method detailed in this paper.

\section{Conclusions} \label{conclusions}

The utilization of polluted WD stars to obtain detailed geochemical characterizations of extrasolar rocky bodies is a powerful tool for elucidating the internal compositions and structures of terrestrial exoplanets. This study analyzes observations of 16 white dwarf stars to understand the oxidation states of the planetary systems at the time and place that the parent bodies of the accreted material formed. We find that most extrasolar rocky bodies formed under oxidizing conditions, comparable to most bodies in the Solar System, including Earth. However, $\thicksim$1/4 of the studied WDs do not yield lower limits in oxygen fugacity, implying more reduced parent bodies may be recorded by a minority of sampled objects.

Intrinsic oxygen fugacity is an important factor that influences fundamental characteristics of a planet, including its volatile budget, its capacity to generate a magnetic field, and its prospects for plate tectonics.  The oxygen fugacities of rocky bodies in the Solar System exhibit a range of \DIW values of $\thicksim$ 7 dex. Earth is indeed oxidized relative to solar, but it is also the most reduced of these rocky bodies excluding Mercury and enstatite chondrites. We conclude that to build an Earth-like planet, a reservoir of relatively reduced rocky material is required. The ability to detect an extrasolar Mercury would confirm that such reservoirs are available.

The difficulty in constraining the oxidation states of these more reduced bodies is discussed and a model for the time-dependent evolution of the apparent oxygen fugacity for a hypothetical reduced body consumed by a WD is investigated. Differences in diffusive fluxes of various elements through the WD envelope yield spurious inferred bulk elemental compositions and oxidation states of the accreting parent bodies under certain conditions. The worst case for biasing against detection of reduced bodies occurs for high effective temperatures. For moderate and low effective temperatures, evidence for relatively reduced parent bodies is preserved under most circumstances for at least several characteristic  lifetimes of the debris disk.  

In general, detailed understanding of both disk lifetimes around WDs and accretion times are important for assessing exoplanetary oxidation states. In order for the current method to {\it unambiguously} detect a low oxygen fugacity parent body in a WD, the white dwarfs would need to be accreting mainly the silicate portion of the body without large mass fractions of metal. In view of the uncertainties, and the importance of characterizing the range of extrasolar rock oxidation states, more attention should be given to the five WDs where lower limits in $x_{\rm FeO}^{\rm rock}$ are consistent with zero in this work. Our calculations suggest that these are candidates for accretion of relatively low oxidation state parent bodies. 

While reduced bodies in the Solar System are less common, the difficulty in constraining them in extrasolar planetary systems will potentially bias our observations against observing them among extrasolar rocks. Therefore, more observations, especially of white dwarfs with debris disks, is the best way to obtain a better statistical characterization of extrasolar rock \DIW values. Strategies for reducing uncertainties in the observations and the extraction of element ratios from the data are also important. We estimate that roughly half the errors in element ratios derive from the raw data themselves and the other half from dispersion in model results for a given star. The strategies may therefore involve longer integration times, in some cases, and the removal of correlated uncertainties from modeled values (Klein et al. in prep.).

It is clear that a better understanding of disk lifetimes is needed in order to draw definitive conclusions about the veracity of element ratios of accreted parent bodies. With this better understanding of the timescales of accretion will come greater confidence in the geochemical parameters derived from polluted WDs. Such evaluations will be the focus of future work. Results will shape future hypotheses for rock formation in the early Solar System and the favored environments for planet formation in general.

\section*{Acknowledgements}
This work was supported by NASA 2XRP grant no. 80NSSC20K0270 to EDY. HES gratefully acknowledges support from the National Aeronautics and Space Administration under grant No. $17~\rm{NAI}18\_~2-0029$ issued through the NExSS Program. We would like to thank the anonymous referee for their comments which improved the manuscript.

\clearpage

\appendix
Here we show the derivation for the exponential decay of a debris disk accreting onto a white dwarf simultaneous with settling of material through the white dwarf atmosphere, as in \cite{Jura2009}. We start with the expression for mass balance of element $z$ in the convective layer of the WD \citep[e.g.][]{Dupuis_1993,RN50}:

\begin{equation}
    \frac{dM_{\rm CV,\it{z}}(t)}{dt} = \dot{M}_z(t) - \frac{M_{\rm CV,\it{z}}(t)}{\tau_z}, 
    \label{massbalance}
\end{equation}

\noindent where $t$ is the elapsed time for accretion, $M_{\rm CV,\it{z}}(t)$ is the mass abundance of element $\it{z}$ in the convection zone, $\dot{M}_z(t)$ is the rate of accretion of element $\it{z}$ onto the star, and $\tau_z$ is the diffusion timescale for element $\it{z}$. The first term on the right side is the rate of accretion of element $\it{z}$ onto the star and the second term is the rate at which element $\it{z}$ leaves the convection zone due to gravitational settling.  We replace  $\dot{M}_z(t)$ with a time variable accretion rate that depends on the mass of the disk at time $t$ and a fixed characteristic disk lifetime $\tau_{\rm disk}$:

\begin{equation}
    \dot{M}_z(t) = \frac{M_{\rm PB,\it{z}}e^{-t/\tau_{\rm disk}}}{\tau_{\rm disk}}.
    \label{Mdot}
\end{equation}

Substituting Equation \ref{Mdot} into Equation \ref{massbalance} gives

\begin{equation}
 \frac{dM_{\rm CV,\it{z}}(t)}{dt} = \frac{M_{\rm PB,\it{z}}e^{\rm -t/\tau_{\rm disk}}}{\tau_{\rm disk}} - \frac{M_{\rm CV,\it{z}}(t)}{\tau_z},
    \label{fullmassbalance}
\end{equation}

\noindent which is a first-order differential equation with the integrating factor $e^{t/\tau_z}$.  Multiplying both sides by the integrating factor yields

\begin{equation}
 e^{t/\tau_z}\frac{dM_{\rm CV,\it{z}}(t)}{dt} + e^{t/\tau_z}\frac{M_{\rm CV,\it{z}}(t)}{\tau_z} =  \frac{M_{\rm PB,\it{z}}e^{\rm -t/\tau_{\rm disk}}e^{t/\tau_z}}{\tau_{\rm disk}}.
    \label{integrationfactor}
\end{equation}

\noindent By the product rule the left-hand side of Equation \ref{integrationfactor} is seen to be $(e^{t/\tau_z}M_{\rm CV,\it{z}})'$, allowing the simplification 

\begin{equation}
 (e^{t/\tau_z}M_{\rm CV,\it{z}})' = \frac{M_{\rm PB,\it{z}}}{\tau_{\rm disk}}e^{\rm {\it t}(1/\tau_z - 1/\tau_{\rm disk})}.
\end{equation}

\noindent Integration of both sides yields

\begin{equation}
 e^{t/\tau_z}M_{\rm CV,\it{z}} = \frac{M_{\rm PB,\it{z}}}{\tau_{\rm disk}} \left [ \frac{e^{\rm (t/\tau_z - t/\tau_{\rm disk})}}{(1/\tau_z - 1/\tau_{\rm disk})} \right ] + c .
\end{equation}

\noindent Solving for $M_{\rm CV,\it{z}}$ 

\begin{equation}
 M_{\rm CV,\it{z}} = \frac{M_{\rm PB,\it{z}}}{\tau_{\rm disk}}\frac{1}{(1/\tau_z - 1/\tau_{\rm disk})} \left [ \frac{e^{\rm ({\it }t/\tau_z - t/\tau_{\rm disk})}}{e^{{\it t}/\tau_z}} \right ] + \frac{c}{e^{t/\tau_z}}
\end{equation}

\noindent and simplifying we obtain

\begin{equation}
 M_{\rm CV,\it{z}} = \frac{M_{\rm PB,\it{z}}}{\tau_{\rm disk}}\frac{1}{(1/\tau_z - 1/\tau_{\rm disk})}e^{- t/\tau_{\rm disk}} + \frac{c}{e^{t/\tau_z}}.
\end{equation}

\noindent Under the initial condition that $M_{\rm CV,\it{z}}$ = 0 at $t = 0$,

\begin{equation}
 0 = \frac{M_{\rm PB,\it{z}}}{\tau_{\rm disk}}\frac{1}{(1/\tau_z - 1/\tau_{\rm disk})}e^{0} + \frac{c}{e^0},
\end{equation}

\noindent and the constant of integration is seen to be

\begin{equation}
 c = (-1)\frac{M_{\rm PB,\it{z}}}{\tau_{\rm disk}}\frac{1}{(1/\tau_z - 1/\tau_{\rm disk})}. 
\end{equation}

\noindent Our solution is therefore

\begin{equation}
\begin{split}
 M_{\rm CV,\it{z}} & = \frac{M_{\rm PB,\it{z}}}{\tau_{\rm disk}}\frac{1}{(1/\tau_z - 1/\tau_{\rm disk})}e^{- t/\tau_{\rm disk}} - \frac{M_{\rm PB,\it{z}}}{\tau_{\rm disk}}\frac{1}{(1/\tau_z - 1/\tau_{\rm disk})}e^{- t/\tau_z}\\
& = \frac{M_{\rm PB,\it{z}}}{\tau_{\rm disk}}\frac{1}{(1/\tau_z - 1/\tau_{\rm disk})} \left [ e^{- t/\tau_{\rm disk}} - e^{- t/\tau_z} \right ]\\
&  = \frac{M_{\rm PB,\it{z}}}{\tau_{\rm disk}}\frac{1}{(1/\tau_z - 1/\tau_{\rm disk})}\frac{\tau_z \tau_{\rm disk}}{\tau_z \tau_{\rm disk}} \left [ e^{- t/\tau_{\rm disk}} - e^{- t/\tau_z} \right ]
   \label{solving}
\end{split}
\end{equation}

\noindent which simplifies to the form shown in \cite{Jura2009}:

\begin{equation}
 M_{\rm CV,\it{z}} = \frac{M_{\rm PB,\it{z}}\tau_z}{\tau_{\rm disk}-\tau_z} \left [ e^{- t/\tau_{\rm disk}} - e^{- t/\tau_z} \right ].
\end{equation}


\end{document}